\documentclass[sigconf]{acmart}
\usepackage{algorithm}
\usepackage{algpseudocode} 
\usepackage{multirow}
\usepackage[normalem]{ulem}
\usepackage{subcaption} 
\useunder{\uline}{\ul}{}

\copyrightyear{2026}
\acmYear{2026}
\setcopyright{cc}
\setcctype{by}
\acmConference[WWW '26] {Proceedings of the ACM Web Conference 2026}{April 13--17, 2026}{Dubai, United Arab Emirates.}
\acmBooktitle{Proceedings of the ACM Web Conference 2026 (WWW '26), April 13--17, 2026, Dubai, United Arab Emirates}
\acmISBN{979-8-4007-2307-0/2026/04}
\acmDOI{10.1145/XXXXXX.XXXXXX}

\AtBeginDocument{%
  }

\begin{document}

\title{PRISM: Personalized Recommendation via Information Synergy Module}


\author{Xinyi Zhang}
\authornotemark[1]
\orcid{0009-0008-1950-3034}
\affiliation{%
  \institution{Imperial College London}
  \city{London}
  \country{UK}}
\email{zxyzxy090588@163.com}
% -----
\author{Yutong Li}
\authornote{Both authors contributed equally to this research.}
\orcid{0009-0008-6496-8850}
\affiliation{%
  \institution{University College London}
  \city{London}
  \country{UK}
}
\email{lyt3612671@163.com}

% -----3
\author{Peijie Sun}
\authornote{Corresponding author.}
\orcid{0000-0001-9733-0521}
\affiliation{%
  \institution{Nanjing University of Posts and Telecommunications}
  \city{Nanjing}
  \country{China}}
\email{sun.hfut@gmail.com}

% -----4
\author{Letian Sha}
\orcid{0009-0007-5547-3728}
\affiliation{%
  \institution{Nanjing University of Posts and Telecommunications}
  \city{Nanjing}
  \country{China}}
\email{ltsha@njupt.edu.cn}

% -----5
\author{Zhongxuan Han}
\orcid{0000-0001-9957-7325}
\affiliation{%
  \institution{Zhejiang University}
  \city{Hangzhou}
  \country{China}
}
\email{zxhan@zju.edu.cn}

\renewcommand{\shortauthors}{Xinyi et al.}

\begin{abstract}

Multimodal sequential recommendation (MSR) leverages diverse item modalities to improve recommendation accuracy, while achieving effective and adaptive fusion remains challenging. Existing MSR models often overlook synergistic information that emerges only through modality combinations. Moreover, they typically assume a fixed importance for different modality interactions across users.
To address these limitations, we propose \textbf{P}ersonalized \textbf{R}ecommend-ation via \textbf{I}nformation \textbf{S}ynergy \textbf{M}odule (PRISM), a plug-and-play framework for sequential recommendation (SR). PRISM explicitly decomposes multimodal information into unique, redundant, and synergistic components through an Interaction Expert Layer and dynamically weights them via an Adaptive Fusion Layer guided by user preferences. This information-theoretic design enables fine-grained disentanglement and personalized fusion of multimodal signals. Extensive experiments on four datasets and three SR backbones demonstrate its effectiveness and versatility. The code is available at~\url{https://github.com/YutongLi2024/PRISM}.
\end{abstract}




\maketitle
\section{Introduction}

Sequential recommendation (SR) is a cornerstone technology for modern Web platforms, designed to tackle information overload by predicting the next item a user will interact with based on their historical behavior sequence~\cite{SASRec, SR_Survey1, SR_Survey2}. With the rapid proliferation of multimodal data across the Web, multimodal information is increasingly incorporated into SR to alleviate data sparsity and cold start issues inherent in traditional ID-based models~\cite{TrustSVD}. Multimodal sequential recommendation (MSR) leverages diverse data sources, such as text, images, and audio, to comprehensively model user preferences and item characteristics, thereby enhancing recommendation performance~\cite{UniSRec, MMSR, MMMLP}.

A core challenge in MSR is modality fusion, which integrates information from heterogeneous modalities into unified representations, critically impacting recommendation effectiveness~\cite{xu2025survey}. Existing methods for learning fused representations primarily focus on aligning modality-shared features or preserving modality-specific features. Most studies~\cite{AlignRec, SEA} adopt self-supervised learning (SSL)~\cite{MMSSL}, particularly contrastive learning, to align representations of the same item across different modalities. As full alignment may dilute the unique information each modality provides, more recent work~\cite{8269806, PAMD, yu2021, DGMRec, liu2022TMM_disentangled, REARM} introduces dedicated encoders or orthogonal learning techniques to decompose features into modality-specific and modality-shared components~\cite{zhang2022}. To match varying user preferences across modalities, adaptive fusion methods~\cite{MMSR, TedRec, HM4SR} leverage attention or perceptual gating mechanisms to dynamically adjust the contribution of each modality, enabling personalized fusion. Despite these advances, MSR methods still face limitations in multimodal fusion, struggling to finely and adaptively disentangle complex multimodal interactions.

\begin{figure}
    \centering
    \includegraphics[width= \linewidth]{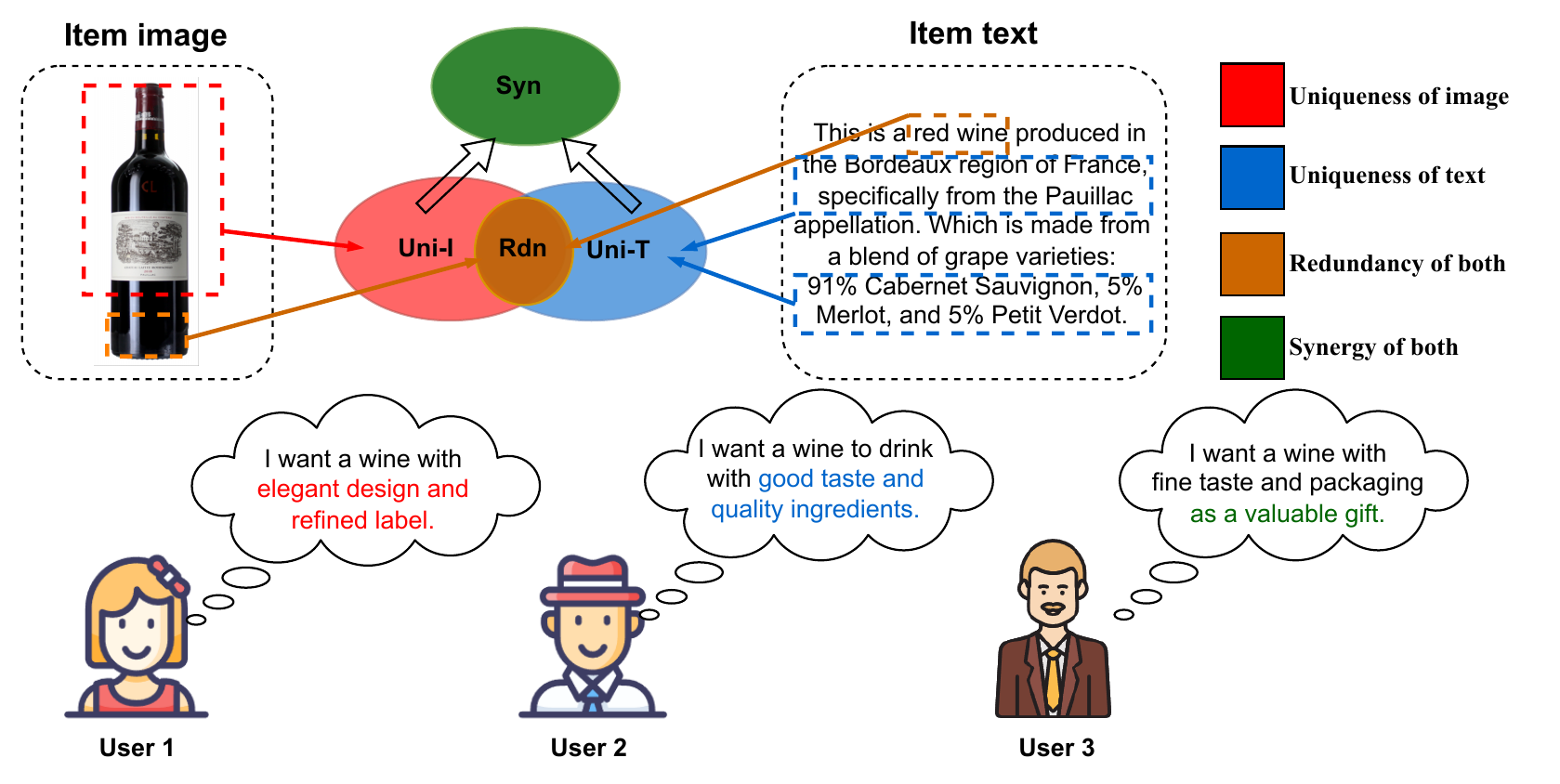}
    \caption{Illustration of multimodal interactions using a red wine example. Higher-level properties, such as the wine’s giftability and collectibility, can only be inferred by synergistically integrating both modalities. The impact of each modality interaction type depends on both user intent and task context.}
    \Description{A diagram showing four types of modality-aware information and three fusion models from baseline to synergy-aware approach.}
    \label{fig:introduction}
\end{figure}

\textbf{Limitation 1: Existing methods fail to capture synergistic information that emerges only from the combination of modalities due to the lack of systematic modeling of multimodal interactions.} Multimodal information can be disentangled into three components: ~\textit{uniqueness} (information specific to one modality), ~\textit{redundancy} (shared information across modalities), and ~\textit{synergy} (incremental information that emerges from the complementary interaction of modalities)~\cite{I2MoE, wollstadt2023, liang2023quantifying, 8269806, liang2023}. Figure~\ref{fig:introduction} illustrates the importance of synergistic information in capturing the complex characteristics of items for accurate recommendation. Specifically, the image modality provides unique cues such as bottle shape, label design, and vintage, while the text modality conveys unique details like production region and grape composition. Information such as color and type appears in both modalities and thus constitutes redundancy. However, to recommend the item as a gift or collectible, the model must leverage the synergy between modalities. Only those wines that exhibit a luxurious appearance as well as a premium origin and composition are suitable for gifting or collection. Existing MSR methods~\cite{DGMRec, REARM} struggle to capture such synergistic signals, as most rely on rigid disentanglement paradigms that either overlook synergy or mistake it for noise. This underscores the urgent need for an information-theoretic approach to model modality interactions and achieve fine-grained disentanglement.

\textbf{Limitation 2: Existing methods typically assume that the relative importance of modality interactions remains consistent across all users and recommendation scenarios, thereby overlooking user-specific preferences in utilizing multimodal information.} In practice, users differ markedly in how they prioritize information~\cite{liu2022TMM_disentangled}. As shown in Figure~\ref{fig:introduction}, some users emphasize unique image cues, valuing aesthetic appeal and product design, while others attend more to unique textual details such as wine quality, taste, and background. Users concerned with giftability or collectibility often rely on higher-level synergistic information. These examples indicate that the influence of each interaction type is highly dependent on user intent and task context. However, existing fusion methods cannot dynamically adjust the weights of different interaction types in a user-aware manner. Although recent approaches~\cite{MGCN, Mentor} recognize that users attend to modalities differently, they typically assign adaptive weights only at a coarse level, treating each modality as an indivisible whole. Such designs obscure which specific information types drive the recommendation outcome, leading to reduced transparency, interpretability, and ultimately user trust. This limitation underscores the need for a dynamic mechanism that captures user interests within the current sequence and adaptively weights information generated from modality interactions.

To overcome these limitations, we propose a general plugin module, \textbf{Personalized Recommendation via Information Synergy Module (PRISM)}. PRISM enables fine-grained disentanglement and personalized fusion of multimodal interaction signals. Designed as \textbf{a plug-and-play component}, PRISM can be seamlessly integrated into existing SR models, guiding parameter optimization through information-theoretic constraints and a reweighting mechanism. (1) To address the first limitation, PRISM introduces an \textbf{interaction expert layer} that leverages interaction-specific losses to train a Mixture-of-Experts (MoE) framework for capturing intricate cross-modal relationships. The uniqueness-preserving loss ensures modality-specific information is retained within fused representations. The redundancy-minimization loss reduces shared information between modalities to yield compact representations. The synergy-capturing loss encourages the model to learn the complementary interplay between modalities. (2) To address the second limitation, PRISM incorporates an \textbf{adaptive fusion layer}, which dynamically assigns importance scores to each interaction expert based on user preferences inferred from historical behavior via an interest-aware mechanism. By explicitly quantifying the contribution of each information component, PRISM enhances transparency and provides interpretable insights into the decision-making process. Extensive experiments across multiple datasets show that PRISM consistently enhances the performance of diverse SR backbones, and when integrated into these backbones, the resulting MSR models can surpass several recently proposed MSR methods, highlighting its effectiveness and flexibility.

Our contributions are summarized as follows:

(1) We propose a novel MoE framework, PRISM, which enables fine-grained decomposition of complex multimodal information. By systematically modeling diverse modality interactions, it addresses the key limitation of MSR, namely the neglect of synergistic information.

(2) We introduce a reweighting mechanism that dynamically assigns personalized importance scores to different interaction types according to user interests, thereby achieving user-aware and interpretable fusion of interaction information.

(3) Extensive experiments across four datasets and three representative backbones demonstrate the effectiveness and generalizability of PRISM.

\section{Methodology}

\begin{figure*}
    \centering
    \includegraphics[width=1\linewidth]{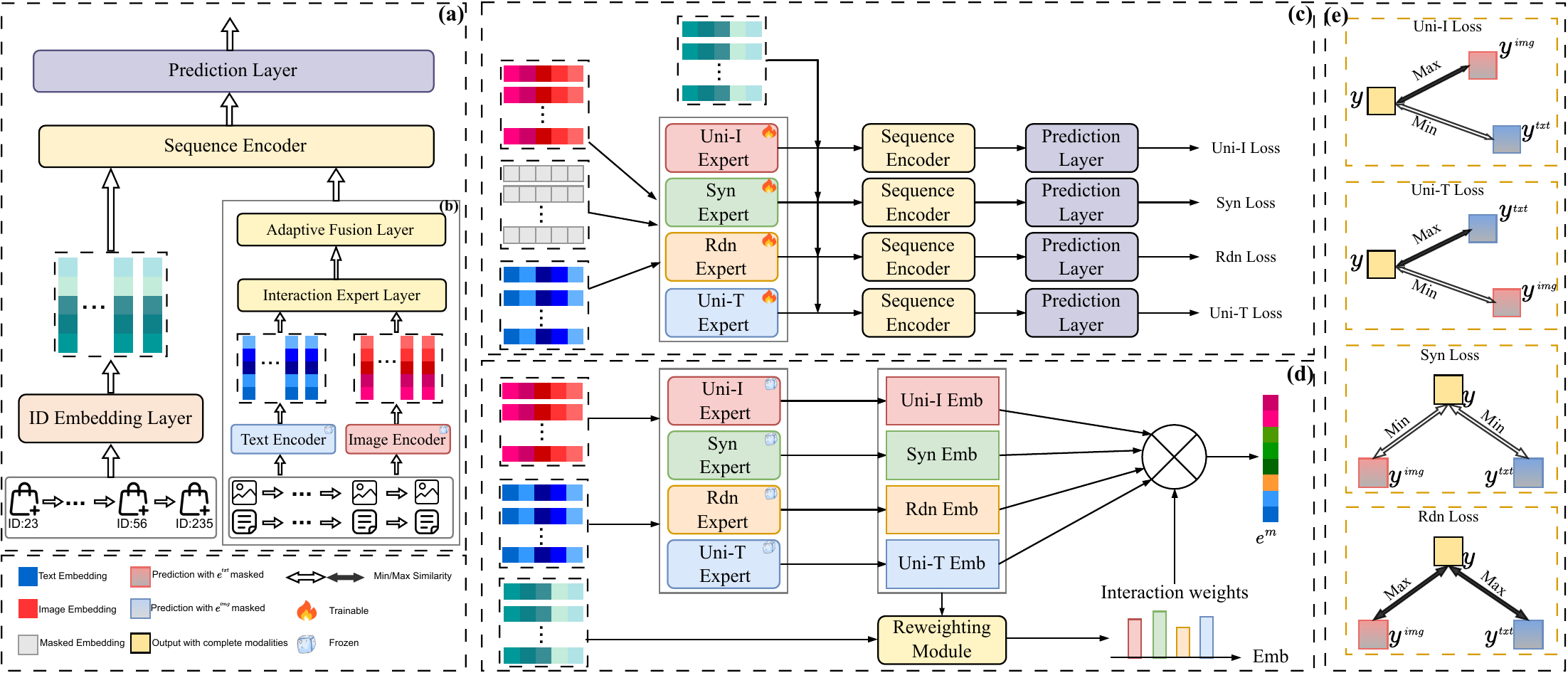}
    \caption{ The overall architecture of the proposed PRISM.}
    \Description{Overview of PRISM, including (a) the overall framework, (b) the Interaction Expert Specialization Layer, and (c) the Adaptive Interaction Fusion Layer.}    
    \label{fig:method}
\end{figure*}

\subsection{Problem Definition}

MSR aims to exploit multimodal information and users’ historical behaviors to generate personalized recommendations for their next interactions. Given a set of users $\mathcal{U}$ and a set of items $\mathcal{I}$, historical interactions can be chronologically organized into sequences. For each user $u \in \mathcal{U}$, we denote the interaction sequence as $\mathcal{S}^u = [x^u_1, x^u_2, \dots, x^u_{|\mathcal{S}^u|}]$, where each item $x^u_i \in \mathcal{I}$ denotes the item that the user interacted with at the $i$-th time step. Following most prior studies~\cite{HM4SR, DGMRec, REARM}, we focus on image and text modalities in our derivations, while the framework can be readily extended to accommodate additional modalities. Formally, each item is represented as $x_i = \{x_i^{id}, x_i^{img}, x_i^{txt}\}$, which incorporates the item ID, image, and textual description. The goal of MSR is to jointly model the sequential dependencies within $\mathcal{S}^u$ and the multimodal features of items to predict the next item a user will interact with.

\subsection{Overview Framework}

In this section, we first introduce the standard MSR pipeline to clearly illustrate how the proposed Personalized Recommendation via Information Synergy Module (PRISM) can be seamlessly integrated into existing SR models. We then elaborate on the technical details of PRISM.

As illustrated in Figure~\ref{fig:method}(a), the interaction sequence of a user is first processed by the ID embedding layer to capture collaborative signals, producing $\mathbf{e}^{id}$ for the sequence~\cite{SASRec,Bert4rec,STOSA}. A positional embedding is then assigned to each item to indicate its relative or absolute position in the sequence. To incorporate multimodal information of items, modality-specific encoders are employed to process $x^{img}$ and $x^{txt}$, and the corresponding latent embeddings are computed as $\mathbf{e}^{img} = \operatorname{imgEmb}(x^{img}), \mathbf{e}^{txt} = \operatorname{textEmb}(x^{txt})$, where $\operatorname{imgEmb}(\cdot)$ and $\operatorname{textEmb}(\cdot)$ are pre-trained CLIP encoders\footnote{\url{https://huggingface.co/sentence-transformers/clip-ViT-B-32}}. Naïve approaches~\cite{PAMD, VBPR} typically fuse $\mathbf{e}^{img}$ and $\mathbf{e}^{txt}$ using simple concatenation or element-wise operations (e.g., summation or averaging). However, such methods fail to explicitly model the heterogeneous interactions between modalities. The resulting fused item representation, which integrates the ID embedding with multimodal signals, is subsequently fed into a sequence encoder, following the architectures of models such as SASRec~\cite{SASRec} and STOSA~\cite{STOSA}, to capture sequential dependencies and user interests.

By replacing rigid multimodal fusion strategies, PRISM enhances the standard SR pipeline with a fine-grained decomposition and personalized fusion of multimodal information. As shown in Figure~\ref{fig:method}(b), PRISM consists of two key components: (1) To address the first limitation, the \textit{Interaction Expert Layer} adopts a MoE framework~\cite{Cai_2025, 10.1145/3219819.3220007} to explicitly disentangle multimodal interactions, where dedicated experts model uniqueness, redundancy, and synergy across modalities. The core innovation of our work lies in the design of tailored loss functions that effectively guide each expert to specialize in its designated type of interaction. (2) To address the second limitation, the \textit{Adaptive Fusion Layer} integrates an interest-aware mechanism to dynamically reweight different interaction types according to user preferences, thereby producing personalized item representations and enhancing interpretability. The optimization of this reweighting module is supervised by the recommendation loss. Through this robust dual-objective optimization, PRISM achieves precise disentanglement of multimodal features while generating fused representations that better align with user preferences. For a clearer understanding of how PRISM is integrated into existing SR backbones, the full pseudocode is provided in Appendix~\ref{appendix:Algorithm}.

\subsection{Interaction Expert Layer}

\subsubsection{Expert Outputs and Training Strategy}

As illustrated in Figures~\ref{fig:method}(c) and (d), the Interaction Expert Layer adopts a MoE framework consisting of four fusion models, each specialized in capturing a specific type of interaction: (1) $E_{\text{uni-i}}$ focuses on unique image information, (2) $E_{\text{uni-t}}$ captures unique textual information, (3) $E_{\text{syn}}$ models synergistic information that emerges from the combination of both modalities, and (4) $E_{\text{rdn}}$ extracts redundant information shared between modalities. Consistent with most existing studies~\cite{HM4SR, M3SRec}, each fusion model is constructed based on a multilayer perceptron (MLP)~\cite{MLP} architecture. This design choice facilitates seamless integration of PRISM into different SR backbones while minimizing additional computational cost and preserving the model’s usability. Each expert functions as an independent fusion model that produces a fused embedding specialized for its respective interaction type:
\begin{equation}
\begin{aligned}
\mathbf{e}^{j} &= E_j(\mathbf{e}^{img}, \mathbf{e}^{txt} ), \quad j \in \{\text{uni-i},\, \text{uni-t},\, \text{syn},\, \text{rdn}\}.
\end{aligned}
\end{equation}

To train initially identical fusion models to specialize in different types of modality interactions, we design an interaction loss that compares unimodal and multimodal recommendation results, following~\cite{yu-etal-2024-mmoe}. First, each interaction expert takes the complete multimodal input and generates a fused representation. This fused representation is then combined with the item ID embedding $\mathbf{e}^{id}$, which is derived from historical interaction sequences, and fed into a sequence encoder $S_j(\cdot)$ and a prediction layer $P_j(\cdot)$ to produce the MSR prediction:
\begin{equation}
\begin{aligned}
\mathbf{y_j} &=P_j(S_j(\mathbf{e}^{id}, E_j(\mathbf{e}^{img},\mathbf{e}^{txt}))), \quad j \in \{\text{uni-i},\, \text{uni-t},\, \text{syn},\, \text{rdn}\}.
\end{aligned}
\end{equation}

Next, we simulate unimodal and perturbed scenarios using a random masking strategy. Specifically, one modality is replaced with a random vector to ensure that the masked modality contributes no residual signal to the recommendation results. For each interaction expert, when the textual embedding is replaced with a random vector $\mathbf{r}$ (sampled from the same space as $\mathbf{e}^{txt}$ and re-sampled at each iteration), the expert generates an image-only prediction $\mathbf{y}^{img}$; conversely, replacing the image embedding $\mathbf{e}^{img}$ with $\mathbf{r}$ yields the text-only prediction $\mathbf{y}^{txt}$:
\begin{align}
\mathbf{y}_{j}^{img} &= P_j\!\big(S_j\!\big(E_j(\mathbf{e}^{img}, \mathbf{r}), \mathbf{e}^{id}\big)\big), 
\qquad j \in \{\text{uni-i}, \text{uni-t}, \text{syn}, \text{rdn}\}, \\
\mathbf{y}_{j}^{txt} &= P_j\!\big(S_j\!\big(E_j(\mathbf{e}^{txt}, \mathbf{r}), \mathbf{e}^{id}\big)\big), 
\qquad j \in \{\text{uni-i}, \text{uni-t}, \text{syn}, \text{rdn}\},
\end{align}
where $\mathbf{r}$ has the same dimensionality as the masked modality and is optionally normalized to match its scale. Compared with zero or mean replacements, stochastic masking more effectively suppresses residual signals and mitigates information leakage, thereby providing a more faithful simulation of unimodal inputs. This is crucial for accurately supervising the learning of uniqueness, redundancy, and synergy. The design is consistent with CoMM~\cite{whattoalign} and is further discussed in Appendix~\ref{appendix:random-vector-justification}. Consequently, each expert produces three prediction signals: $\mathbf{y}$ for the full multimodal input, $\mathbf{y}^{img}$ for the text-masked condition, and $\mathbf{y}^{txt}$ for the image-masked condition.

\subsubsection{Interaction Loss}

As shown in Figure~\ref{fig:method}(e), we define expert-specific interaction losses that realize Partial Information Decomposition (PID)~\cite{wollstadt2023, liang2023quantifying, 8269806, liang2023} through anchor–positive–negative configurations and cosine similarity computed over $(\mathbf{y}, \mathbf{y}^{img}, \mathbf{y}^{txt})$. For all experts, the prediction generated from the complete multimodal input, $\mathbf{y}$, serves as the anchor.

\textbf{Uniqueness loss.}
Uniqueness experts aim to extract modality-specific information while suppressing signals from the other modality. For the image uniqueness expert $E_{\text{uni-i}}$, the output $\mathbf{y}^{img}$ obtained when only the image embedding $\mathbf{e}^{img}$ is preserved is treated as the positive, since it retains the unique image cues. In contrast, $\mathbf{y}^{txt}$, which results from masking $\mathbf{e}^{img}$, serves as the negative, as this setting discards the very information the expert is expected to capture. The objective is to encourage $\mathbf{y}$ to be highly similar to $\mathbf{y}^{img}$ while dissimilar to $\mathbf{y}^{txt}$. We adopt the triplet loss~\cite{tripletloss}, defined as $\mathrm{Triplet}(a,p,n) = \max\!\left(0, m + d(a,p) - d(a,n)\right)$ with $d(a,b) = 1-\mathrm{CosSim}(a,b)$, where $m>0$ is a margin hyperparameter:
\begin{equation}
\mathcal{L}_{\text{uni-i}} = \texttt{TripletLoss}\big(\mathbf{y},\, \mathbf{y}^{img},\, \mathbf{y}^{txt}\big).
\end{equation}

The design is symmetric across modalities. For the textual uniqueness expert $E_{\text{uni-t}}$, $\mathbf{y}^{txt}$, obtained by preserving $\mathbf{e}^{txt}$, is treated as the positive sample, whereas $\mathbf{y}^{img}$, obtained by masking $\mathbf{e}^{txt}$, serves as the negative sample. The objective is to make $\mathbf{y}$ closely aligned with $\mathbf{y}^{txt}$ while remaining clearly separated from $\mathbf{y}^{img}$, as $E_{\text{uni-t}}$ is designed to capture the unique information contained in $\mathbf{e}^{txt}$:
\begin{equation}
\mathcal{L}_{\text{uni-t}} = \texttt{TripletLoss}\big(\mathbf{y},\, \mathbf{y}^{txt},\, \mathbf{y}^{img}\big).
\end{equation}

\textbf{Synergy loss.}
The synergy expert is designed to capture emergent information that arises exclusively from the joint presence of both modalities. When either modality is masked, this synergistic interaction collapses; hence, both $\mathbf{y}^{img}$ and $\mathbf{y}^{txt}$ are treated as negatives. To enforce this property, we minimize the similarity between the multimodal prediction $\mathbf{y}$ and the perturbed outputs $\mathbf{y}^{img}$ and $\mathbf{y}^{txt}$. We employ cosine similarity~\cite{Cossim}, defined as $\mathrm{CosSim}(a,b)=\tfrac{a^\top b}{|a|_2 |b|_2}$, which measures the directional alignment between two vectors independent of their magnitude. The synergy loss is formulated as:
\begin{equation}
\mathcal{L}_{\text{syn}} = \tfrac{1}{2}\Big[ \texttt{CosSim}(\mathbf{y}, \mathbf{y}^{img}) + \texttt{CosSim}(\mathbf{y}, \mathbf{y}^{txt}) \Big],
\end{equation}
This encourages $\mathbf{y}$ to diverge from both unimodal perturbations, ensuring that the synergy expert focuses on the complementary signals jointly encoded by $\mathbf{e}^{img}$ and $\mathbf{e}^{txt}$.

\textbf{Redundancy loss.}  
Redundancy refers to information consistently present in both modalities. In this case, masking one modality should not fundamentally alter the prediction. Therefore, $\mathbf{y}^{img}$ and $\mathbf{y}^{txt}$ are both positives relative to $\mathbf{y}$. Using the same cosine similarity definition, the loss encourages high similarity among all three predictions, so that $E_{\text{rdn}}$ focuses on redundantly encoded features in $\mathbf{e}^{img}$ and $\mathbf{e}^{txt}$:
\begin{equation}
\mathcal{L}_{\text{rdn}} = 1 - \tfrac{1}{2}\Big[ \texttt{CosSim}(\mathbf{y}, \mathbf{y}^{img}) + \texttt{CosSim}(\mathbf{y}, \mathbf{y}^{txt}) \Big].
\end{equation}

Finally, the overall interaction loss integrates the four objectives with separate tunable weights:
\begin{equation}
\mathcal{L_{\text{exp}}} = \lambda_{\text{uni-i}} \cdot \mathcal{L}_{\text{uni-i}} + \lambda_{\text{uni-t}} \cdot \mathcal{L}_{\text{uni-t}} + \lambda_{\text{syn}} \cdot \mathcal{L}_{\text{syn}} + \lambda_{\text{rdn}} \cdot \mathcal{L}_{\text{rdn}},
\end{equation}
where $\lambda_{\text{uni-i}}$, $\lambda_{\text{uni-t}}$, $\lambda_{\text{syn}}$, and $\lambda_{\text{rdn}}$ are hyperparameters that control the relative importance of image uniqueness, textual uniqueness, synergy, and redundancy, respectively. This formulation provides finer-grained control compared to an averaged weighting, ensuring that each type of interaction can be emphasized or down-weighted depending on task requirements. The theoretical connection between the proposed interaction loss and the four components of PID is elaborated in Appendix~\ref{appendix:connection-loss-and-pid}.

\subsection{Adaptive Fusion Layer}

While the Interaction Expert Layer outputs interaction-specific fused embeddings $\mathbf{e}^{uni-i}$, $\mathbf{e}^{\text{uni-t}}$, $\mathbf{e}^{\text{syn}}$, and $\mathbf{e}^{\text{rdn}}$, their relative contributions to recommendation depend on user preferences and recommendation contexts. To address this limitation, we design an adaptive reweighting mechanism that dynamically adjusts the importance of each interaction type, enabling the model to generate informative multimodal fusion tailored to each user sequence.

As shown in Figure~\ref{fig:method}(d), we take the interaction-specific embeddings and integrate them with the ID embedding $\mathbf{e}^{id}$, which encodes sequential user–item interactions and serves as the basic representation of user preferences, following SASRec~\cite{SASRec} and STOSA~\cite{STOSA}. The concatenated embeddings are then passed through a reweighting module $\mathrm{W}$. We instantiate $\mathrm{W}$ as a MLP~\cite{MLP}, which offers a lightweight yet expressive means of modeling non-linear dependencies among embeddings, without the added complexity and overhead of attention-based architectures~\cite{Transformer}:
\begin{equation}
w^j = \mathrm{W}\!\big(\mathbf{e}^{\text{uni-i}},\, \mathbf{e}^{\text{uni-t}},\, \mathbf{e}^{\text{syn}},\, \mathbf{e}^{\text{rdn}},\, \mathbf{e}^{id}\big), j \in \{\text{uni-i}, \text{uni-t}, \text{syn}, \text{rdn}\},
\end{equation}
where $w^j$ denotes the importance score assigned to a specific interaction type, indicating its contribution to the preference representation of the user within the current sequence. In contrast to existing fusion strategies that treat each modality as a whole and only adjust modality-level weights, our reweighting module allows the model to flexibly prioritize modality-unique, synergistic, or redundant information based on the recommendation context.

The final multimodal representation is obtained by weighting each interaction embedding with its corresponding learned coefficient and aggregating the results:
\begin{equation}
\mathbf{e}^m = \sum_{j} w^j \, \mathbf{e}^{\,j}, \quad j \in \{\text{uni-i},\, \text{uni-t},\, \text{syn},\, \text{rdn}\}.
\end{equation}

This reweighted fusion mechanism offers two key advantages. First, it enables personalized fusion by adaptively emphasizing the most relevant interaction types for each user, thereby enhancing recommendation accuracy. Second, the explicit weights $w^j$ provide interpretability by quantifying the relative contributions of uniqueness, synergy, and redundancy, allowing direct analysis of which aspects of multimodal information influence user preferences. We further demonstrate this interpretability in Section~\ref{Case Study}, where visualizations of the learned weight distributions reveal how the model dynamically adapts to varying recommendation contexts.

\subsection{Prediction and Optimization}

As a plug-and-play module, PRISM is designed for seamless integration with a wide range of SR backbones. Its overall training objective comprises two components: the native recommendation loss of the backbone, $\mathcal{L_{\text{rec}}}$, and the multimodal interaction loss, $\mathcal{L_{\text{exp}}}$. To ensure maximum compatibility, PRISM directly inherits the original loss of the backbone, $\mathcal{L_{\text{rec}}}$, allowing it to adapt to the diverse training objectives used in existing SR models. For instance, $\mathcal{L_{\text{rec}}}$ can be the Binary Cross-Entropy (BCE)~\cite{BCE}, which is widely adopted in models such as SASRec~\cite{SASRec}. Alternatively, it can be the Bayesian Personalized Ranking (BPR)~\cite{BPR}, which serves as the core objective in models like STOSA~\cite{STOSA}.

By adopting the native objective of the SR backbone without modification, PRISM can be readily applied to a wide spectrum of current and future architectures. The final training objective integrates both loss components as follows:
\begin{equation}
\mathcal{L} = \mathcal{L_{\text{rec}}} + \mathcal{L_{\text{exp}}}.
\end{equation}
\section{Experiment}

In this section, we investigate the following research questions:
\textbf{RQ1:} How does PRISM perform relative to the baseline methods?
\textbf{RQ2:} What is the contribution of each component to the overall PRISM architecture?
\textbf{RQ3:} How sensitive is PRISM to different hyperparameter settings?
\textbf{RQ4:} Can PRISM effectively capture synergistic interactions and user-guided fusion to improve recommendation quality?
\textbf{RQ5:} What are the time and space complexities of PRISM?

\subsection{Experimental Setup}

\textbf{Datasets.} We evaluate PRISM on four real-world datasets: Amazon\footnote{\url{https://snap.stanford.edu/data/amazon/productGraph}} Home, Beauty, and Sports, as well as Yelp\footnote{\url{https://business.yelp.com/data/resources/open-dataset}}. The Amazon datasets capture users’ purchasing behaviors and product reviews across diverse domains, and are widely used benchmarks for SR~\cite{SASRec, STOSA, HM4SR, M3SRec}. The Yelp dataset, collected from a local business review platform, focuses on user feedback for restaurants and other services, and is commonly adopted in recommendation research~\cite{Yelp1, Yelp2}. All datasets are preprocessed using the standard 5-core setting~\cite{SASRec, STOSA, Yelp1, Yelp2}, ensuring that each user and item has at least five interactions.
Detailed statistics are summarized in Table~\ref{tab:Statistics}.

\begin{table}[!ht]
\centering
\caption{Datasets Statistics}
\resizebox{0.45\textwidth}{!}{%
\begin{tabular}{@{}cccccc@{}}
\toprule
Dataset & \multicolumn{1}{c}{\#users} & \multicolumn{1}{c}{\#items} & \multicolumn{1}{c}{\#interactions} & \multicolumn{1}{c}{density} & \multicolumn{1}{c} {avg.length} \\ 
\midrule
Home & 66,519 & 28,238 & 551,682 & 0.03\% & 8.3 \\
Beauty & 22,363 & 12,102 & 198,502 & 0.07\% & 8.9 \\
Sports& 35,598& 18,358& 296,337& 0.05\%& 8.3\\
Yelp & 287,116 & 148,523 & 4,392,169 & 0.01\% & 15.1\\
\bottomrule
\end{tabular}%
}
\label{tab:Statistics}
\end{table}

\textbf{Evaluation Settings.}
To ensure fair and rigorous evaluation, we follow the standard leave-one-out protocol for dataset splitting~\cite{SASRec, STOSA}. For each user sequence, the last item is held out for testing, the second-to-last for validation, and all preceding items are used for training. We report two widely used metrics, Recall@K (R@K)~\cite{LightGCN, STOSA, TedRec, DGMRec} and NDCG@K (N@K)~\cite{SASRec, Bert4rec, MMMLP, HM4SR}, with $K \in \{10, 20\}$. Higher values of R@K and N@K indicate more accurate recommendation results.

\textbf{Baseline Methods.}
To evaluate the effectiveness and versatility of PRISM, we compare it against representative and competitive baselines from three categories:
(1) \textbf{Traditional Recommendation:} Methods relying on item co-occurrence patterns, including SASRec~\cite{SASRec}, BERT4Rec~\cite{Bert4rec}, LightGCN~\cite{LightGCN}, STOSA~\cite{STOSA}, DiffuRec~\cite{Diffurec_TOIS}, and InDiRec~\cite{InDiRec};
(2) \textbf{Multimodal Recommendation:} Methods that integrate additional modalities such as images and texts, including VBPR~\cite{VBPR}, UniSRec~\cite{UniSRec}, MMMLP~\cite{MMMLP}, MMSBR~\cite{MMSBR}, TedRec~\cite{TedRec}, and HM4SR~\cite{HM4SR};
(3) \textbf{Multimodal Recommendation (Focused on Uniqueness and Redundancy):} Methods that disentangle modality-unique and modality-shared information via contrastive learning, orthogonality, or graph modeling, but do not model synergistic information. Representative examples include PAMD~\cite{PAMD}, DGMRec~\cite{DGMRec}, and REARM~\cite{REARM}.

\textbf{Implementation Details.}
All models are implemented in PyTorch, building upon the widely used open-source libraries named RecBole\footnote{\url{https://github.com/RUCAIBox/RecBole2.0}}~\cite{RecBole}. To ensure fair comparison, baseline models are either configured with the best hyperparameters reported in their original papers or tuned via grid search for datasets not covered previously. For PRISM, hyperparameters are optimized on the validation set using grid search. Four interaction-related coefficients, which are $\lambda_{\text{uni-i}}$, $\lambda_{\text{uni-t}}$, $\lambda_{\text{syn}}$, and $\lambda_{\text{rdn}}$, control the relative importance of image/text uniqueness, synergy, and redundancy, each searched over $\{0.01, 0.05, 0.1, 0.2, 0.5, 1.0\}$.
The final values are reported in Section~\ref{sec:hyperparam}. All experiments are repeated five times with different random seeds, and the average performance is reported. Statistical significance is verified via paired t-tests with $p < 0.05$. All experiments are conducted on an NVIDIA RTX 4090 GPU.

\begin{table*}[t]
    \small
    \setlength\tabcolsep{2pt}
    \caption{Performance comparison with three groups of baselines. The best results are highlighted in \textbf{bold}, and the second-best results are \underline{underlined}. \emph{vs. Vanilla} denotes the relative improvement over the vanilla model, while \emph{vs. Best} indicates the relative improvement over the strongest baseline (in percentage). All results are averaged over 5 runs per dataset for statistical robustness and are statistically significant at \(p < 0.05\).}
    \label{tab:compare}
    \vspace{-5pt}
    \resizebox{0.96\linewidth}{!}{
        \begin{tabular}{c|cccc|cccc|cccc|cccc}
        \hline
        \makebox[0.15\linewidth]{Datasets}        & \multicolumn{4}{c|}{Home} & \multicolumn{4}{c|}{Beauty}  & \multicolumn{4}{c|}{Sports}    & \multicolumn{4}{c}{Yelp}                                                                             \\ \hline
        Metric                                    & {R@10}& {R@20}& {N@10}&{N@20}                      & {R@10}& {R@20}& {N@10}& {N@20}                      
        &{R@10}& {R@20}& {N@10}& {N@20}                       & {R@10}& {R@20}& {N@10}& {N@20}  \\ \hline
        
        SASRec (ICDM'18)                          & 0.0168& 0.0249& 0.0081& 0.0099                     & 0.0534& 0.0839& 0.0247& 0.0332                      
        &0.0336& 0.0505& 0.0175& 0.0218           & 0.0233& 0.0391& 0.0123& 0.0152  \\
        BERT4Rec (CIKM'19)                        & 0.0156& 0.0238& 0.0073& 0.0101                     & 0.0545& 0.0852& 0.0254& 0.0351                      
        &0.0342& 0.0512& 0.0172& 0.0216           & 0.0245& 0.0411& 0.0119& 0.0162  \\ 
        LightGCN (SIGIR'20)                       & 0.0161& 0.0243& 0.0077& 0.0103                     & 0.0549& 0.0861& 0.0258& 0.0355                      
        &0.0366& 0.0535& 0.0203& 0.0242           & 0.0236& 0.0414& 0.0123& 0.0166  \\
        STOSA (WWW'22)                            & 0.0169& 0.0264& 0.0098& 0.0113                     & 0.0648& 0.0941& 0.0339& 0.0385                      
        &0.0389& 0.0560& 0.0220& 0.0264           & 0.0238& 0.0424& 0.0128& 0.0161  \\
        DiffuRec (TOIS'23)                        & 0.0278& 0.0369& 0.0185& 0.0208                     & 0.0854& 0.1061& 0.0548& 0.0642                     
        &0.0481& 0.0721& 0.0305& 0.0361           & 0.0324& 0.0499& 0.0221& 0.0265  \\
        InDiRec (SIGIR'25)                        & 0.0330& 0.0452& 0.0202& 0.0232                     & 0.0941& 0.1286&0.0573&0.0660                      
        &0.0522& 0.0752& 0.0309& 0.0367           & 0.0372& 0.0551& 0.0262& 0.0301  \\ \hline
        
        VBPR (AAAI'16)                            & 0.0159& 0.0256& 0.0074& 0.0095                     & 0.0551& 0.0842& 0.0252& 0.0337                      
        &0.0321& 0.0492& 0.0159& 0.0195           & 0.0237& 0.0434& 0.0123& 0.0163  \\
        UniSRec (KDD'22)                          & 0.0198& 0.0286& 0.0121& 0.0129                     & 0.0661& 0.0988& 0.0386& 0.0439                      
        &0.0361& 0.0544&0.0199&0.0223             & 0.0246& 0.0442& 0.0137& 0.0167  \\
        MMMLP (WWW'23)                            & 0.0237& 0.0327& 0.0145& 0.0149         
            &0.0725& 0.1056&0.0422&0.0501         
        & 0.0381& 0.0567& 0.0214& 0.0233          &0.0279& 0.0483&0.0146&0.0161   \\
        MMSBR (TKDE'23)                           & 0.0238& 0.0331& 0.0142& 0.0152                    & 0.0719& 0.1062&0.0406&0.0474         
        & 0.0393& 0.0579& 0.0235& 0.0245          &0.0255& 0.0464&0.0139& 0.0166  \\
        TedRec (CIKM'24)                           & 0.0305& 0.0431& 0.0174& 0.0208                    &0.0927& 0.1188& 0.0527& 0.0631        
        & 0.0501& 0.0725& 0.0308 & 0.0369            &0.0351& 0.0504& 0.0215& 0.0261  \\
        HM4SR (WWW'25)                             & {\ul0.0333}& {\ul0.0461}& 0.0191& 0.0227                    & 0.0930& 0.1299& 0.0565& 0.0651       
        & {\ul0.0528}& {\ul0.0761}& 0.0301& 0.0359          & {\ul0.0381}& {\ul0.0573}& 0.0249& 0.0296  \\ \hline
        
        PAMD (WWW'22)                              &0.0245&0.0340&0.0148&0.0156                 
        & 0.0733& 0.1071&0.0418&0.0485             
        &0.0390&0.0574&0.0221&0.0241               &0.0267& 0.0471&0.0142&0.0164   \\
        DGMRec (SIGIR'25)                           & 0.0327& 0.0455&{\ul0.0210}&{\ul0.0243}                    &0.0943&0.1301&{\ul0.0579}&{\ul0.0665}        
        &0.0513& 0.0748& 0.0312 & 0.0371            &0.0371& 0.0548& 0.0269& 0.0310  \\
        REARM (MM'25)                             & 0.0331& 0.0458& 0.0206& 0.0231                      &{\ul0.0954}&{\ul0.1311}&0.0574&0.0659       
        &0.0520& 0.0755&{\ul0.0319}&{\ul0.0382}     &0.0380& 0.0560&{\ul0.0275}&{\ul0.0317}\\ \hline
        
        PRISM+SASRec         & 0.0221& 0.0309& 0.0116& 0.0139          & 0.0681& 0.1051& 0.0343& 0.0457       
        & 0.0406& 0.0613& 0.0227& 0.0291                       & 0.0271& 0.0448& 0.0141& 0.0173  \\
        \emph{vs. Vanilla}              & 25.60\%& 24.10\%& 43.21\%& 40.40\%                    & 27.53\%& 25.27\%& 38.87\%& 37.65\%                     
        & 20.83\%& 21.39\%& 29.71\%& 33.49\%                    & 16.31\%& 14.58\%& 14.63\%& 13.82\% \\
        PRISM+STOSA                  & 0.0261& 0.0398& 0.0168& 0.0193                     
        & 0.0762& 0.1083& 0.0481& 0.0549                      
        & 0.0433& 0.0621& 0.0258& 0.0301                       & 0.0285& 0.0511& 0.0161& 0.0195  \\
        \emph{vs. Vanilla}              & 54.44\%& 50.76\%& 71.43\%& 70.80\%                    & 17.59\%& 15.09\%& 41.89\%& 42.60\%                     
        &11.31\%& 10.89\%& 17.27\%& 14.02\%                    & 19.75\%& 20.52\%& 25.78\%& 21.12\%     \\
        PRISM+InDiRec                    & \textbf{0.0364}& \textbf{0.0501}& \textbf{0.0235}& \textbf{0.0272}
        &\textbf{0.1012}& \textbf{0.1388}& \textbf{0.0618}& \textbf{0.0711}             & \textbf{0.0557}& \textbf{0.0809}& \textbf{0.0337}& \textbf{0.0401}              & \textbf{0.0422}& \textbf{0.0622}& \textbf{0.0305}& \textbf{0.0352}  \\
        \emph{vs. Vanilla}              & 10.30\%& 10.84\%& 16.34\%& 17.24\%                  & 7.55\%& 7.93\%& 7.85\%& 7.73\%                     
        & 6.70\%& 7.58\%& 9.06\%& 9.26\%                    & 13.44\%& 12.89\%& 16.41\%& 16.94\% \\ \hline
        \emph{vs. Best}                & 9.31\%& 8.68\%& 11.90\%& 11.93\%                  & 6.08\%& 5.87\%& 6.74\%& 6.92\%                   
        &5.49\%& 6.31\%& 5.64\%& 4.97\%                     & 10.76\%& 8.55\%& 10.91\%& 11.04\% \\ \hline
        \end{tabular}
        }
    \vspace{-8pt}
\end{table*}

\subsection{Overall Performance Comparison (RQ1)}

This subsection presents a comprehensive comparison to evaluate the performance of PRISM. As shown in Table~\ref{tab:compare}, PRISM consistently improves the performance of SR models across all datasets and evaluation metrics. (1) Compared with ID-based SR models, PRISM-equipped variants achieve substantial performance gains. This clear margin demonstrates the advantage of incorporating multimodal information into sequential modeling. (2) Compared with representative multimodal baselines such as TedRec~\cite{TedRec} and HM4SR~\cite{HM4SR}, PRISM-equipped models, particularly PRISM+InDiRec, achieve superior results on all datasets. For instance, on the Home dataset, PRISM+InDiRec attains 0.0364 in R@10 and 0.0235 in N@10, surpassing HM4SR (0.0333 and 0.0191, respectively). These improvements highlight the effectiveness of the user-guided fusion in PRISM, which dynamically adjusts the importance of different modality interactions based on user preferences, overcoming the rigidity of fixed-weight fusion strategies. (3) Compared with disentanglement approaches such as DGMRec~\cite{DGMRec} and REARM~\cite{REARM}, PRISM also demonstrates clear superiority. On the Yelp dataset, PRISM+InDiRec outperforms REARM, which is the strongest baseline in this group, by 10.91\% in N@10 and 11.04\% in N@20. This confirms that explicitly modeling modality synergy yields additional performance gains. (4) PRISM serves as a plug-and-play module compatible with various SR backbones, including SASRec~\cite{SASRec}, STOSA~\cite{STOSA}, and InDiRec~\cite{InDiRec}. The consistent improvements over the vanilla counterparts, such as the 71.43\% increase in N@10 when integrated with STOSA on the Home dataset, further validate PRISM’s versatility and broad applicability across architectures.

\subsection{Ablation Study (RQ2)}\label{ablation}

To evaluate the contribution of each component in PRISM, we conduct ablation experiments on the Yelp dataset using PRISM+InDiRec as the target model. Specifically, we compare PRISM+InDiRec with the following variants: (\textit{i}) \textit{w/o} $E_{\text{uni-i}}$, which removes the image uniqueness expert; (\textit{ii}) \textit{w/o} $E_{\text{uni-t}}$, which discards the textual uniqueness expert; (\textit{iii}) \textit{w/o} $E_{\text{syn}}$, which eliminates the synergy expert; (\textit{iv}) \textit{w/o} $E_{\text{rdn}}$, which excludes the redundancy expert; and (\textit{v}) \textit{w/o} $AFL$, which disables the Adaptive Fusion Layer while keeping all other components intact. As shown in Table~\ref{table:ablation}, the full PRISM+InDiRec model achieves the best performance across all evaluation metrics, demonstrating the effectiveness of jointly modeling uniqueness, redundancy, and synergy interactions with adaptive fusion. Among all variants, removing the textual uniqueness expert (\textit{w/o} $E_{\text{uni-t}}$) leads to the largest performance drop, indicating that text-specific cues play a crucial role in capturing user intent in the Yelp dataset. The degradation caused by removing the image uniqueness expert (\textit{w/o} $E_{\text{uni-i}}$) further confirms the importance of preserving modality-specific image signals. Comparing these two variants suggests that textual information is generally more informative than visual features, as text more directly conveys item semantics. The removal of the synergy expert (\textit{w/o} $E_{\text{syn}}$) also results in a notable decline, highlighting that emergent cross-modal interactions cannot be replaced by unimodal signals alone and are indispensable for capturing higher-level semantics. In contrast, excluding the redundancy expert (\textit{w/o} $E_{\text{rdn}}$) causes the smallest degradation, implying that shared redundant information contributes less to modeling user preferences. Finally, disabling the Adaptive Fusion Layer (\textit{w/o} $AFL$) consistently reduces performance across all metrics, underscoring the necessity of dynamically weighting interaction types according to user behavior. Overall, although the degree of degradation varies across ablations, the consistent performance decline demonstrates that each component of PRISM contributes complementary benefits, and their joint integration is essential for achieving robust multimodal recommendation performance.

\begin{table}[!t]
    \caption{Ablation results of different PRISM+InDiRec variants. “\textit{w/o}” denotes “without.” All results are averaged over five runs on the Yelp dataset for statistical robustness and are statistically significant with \(p < 0.05\).}
    \label{table:ablation}
    \centering
    \begin{tabular}{lcccc}
        \toprule
        \textbf{Variant}&   \textbf{R@10} & \textbf{R@20} & \textbf{N@10}& \textbf{N@20} \\
        \midrule
        PRISM \textit{(original)} & \textbf{0.0422} & \textbf{0.0622} & \textbf{0.0305} & \textbf{0.0352}\\
        \textit{w/o} $E_{\text{uni-i}}$ &  0.0376 & 0.0582 & 0.0271 & 0.0320\\
        \textit{w/o} $E_{\text{uni-t}}$ &  0.0361 & 0.0570 & 0.0257 & 0.0300\\
        \textit{w/o} $E_{\text{syn}}$ &  0.0381 & 0.0571 & 0.0269 & 0.0301\\
        \textit{w/o} $E_{\text{rdn}}$ &  0.0391 & 0.0594 & 0.0277 & 0.0325\\
        \textit{w/o} $AFL$ & 0.0386 & 0.0589 & 0.0278 & 0.0327\\
        \bottomrule
    \end{tabular}
    
\end{table}

\subsection{Hyperparameter Analysis (RQ3)}\label{sec:hyperparam}

\begin{figure}
    \centering
    \includegraphics[width=1\linewidth]{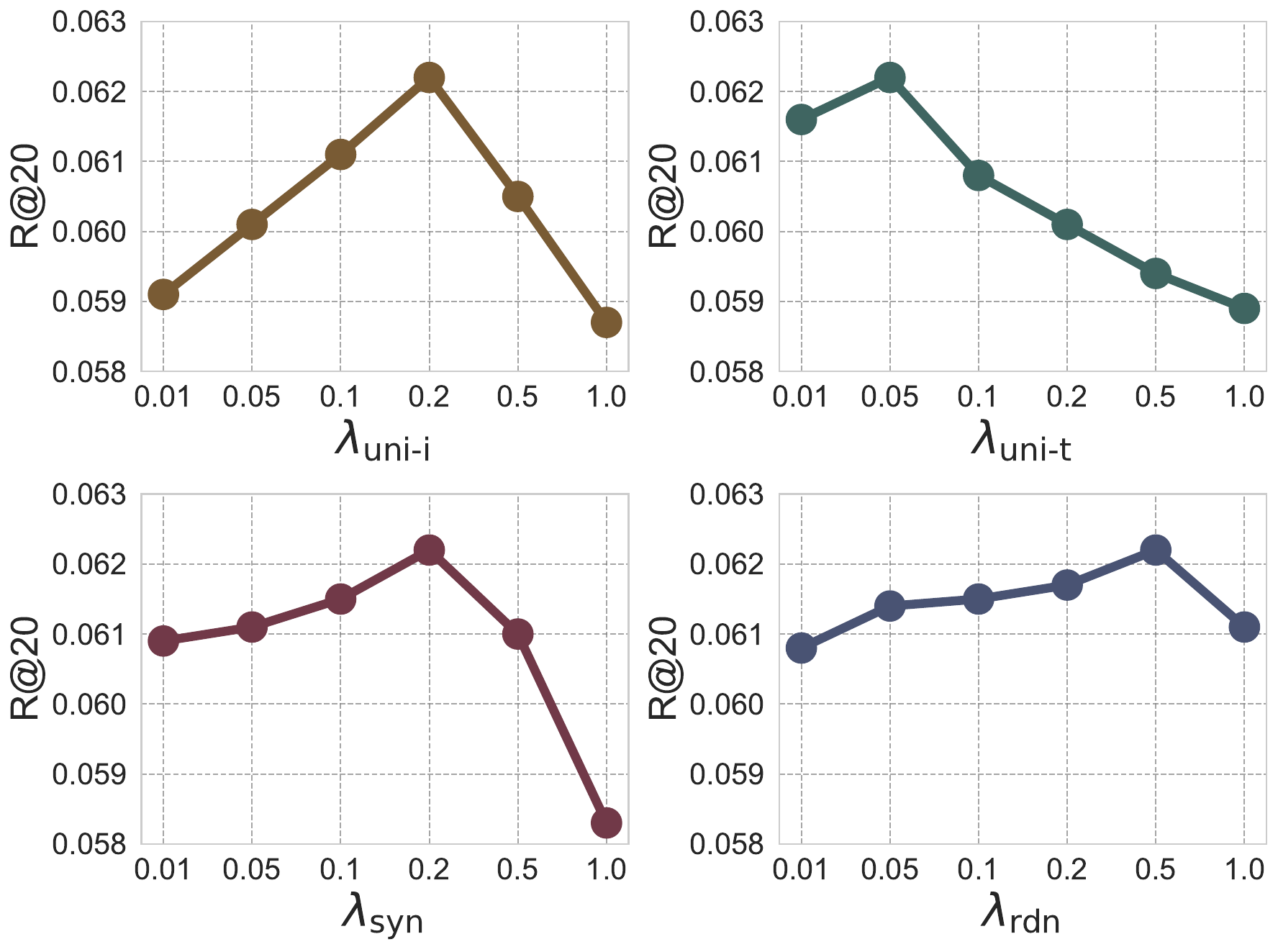}
    \caption{Performance comparison of PRISM+InDiRec with different hyperparameter settings on the Yelp dataset.}
    \Description{A Figure showing Hyperparameter comparison.}
    \label{fig:hyper}
\end{figure}

This subsection examines the sensitivity of PRISM to the four interaction-loss coefficients, $\lambda_{\text{uni-i}}$, $\lambda_{\text{uni-t}}$, $\lambda_{\text{syn}}$, and $\lambda_{\text{rdn}}$, using the PRISM+InDiRec model on the Yelp dataset, as illustrated in Figure~\ref{fig:hyper}. We vary each coefficient within $\{0.01, 0.05, 0.1, 0.2, 0.5, 1.0\}$ while keeping the others fixed at their default values. Overall, the coefficients for uniqueness and synergy losses require careful calibration, whereas the redundancy coefficient exhibits greater tolerance to variation. For the image uniqueness coefficient $\lambda_{\text{uni-i}}$, performance gradually improves as the coefficient increases from 0.01 to 0.2, peaking at 0.0622. This indicates that moderate emphasis on image-specific cues, such as object shape or packaging, enhances user-intent modeling; however, excessive weighting causes a sharp decline due to the suppression of textual uniqueness and synergistic information. The textual uniqueness coefficient $\lambda_{\text{uni-t}}$ achieves the best result at 0.05, beyond which R@20 declines steadily to 0.0589 at 1.0. This pattern suggests that while moderate textual emphasis strengthens semantic grounding, overemphasis leads to modality dominance, overshadowing other signals. For the synergy coefficient $\lambda_{\text{syn}}$, the performance curve exhibits a convex shape and reaches its maximum of 0.0622 at 0.2, confirming that synergy is crucial for capturing high-level semantics inaccessible to single modalities. Nevertheless, overly large values destabilize training by excessively penalizing unimodal predictions. In contrast, the redundancy coefficient $\lambda_{\text{rdn}}$ shows a relatively flat trend, with the best performance of 0.0622 at 0.5 and only marginal degradation beyond that point. This implies that redundancy primarily acts as a stabilizer in multimodal representation learning and is less sensitive to hyperparameter variation. In summary, uniqueness and synergy require moderate weighting (0.05--0.2) to achieve optimal balance, while redundancy remains robust across a broader range of values.

\subsection{Case Study (RQ4)}\label{Case Study}

\begin{figure*}[!ht]
    \centering
    \includegraphics[width=0.95\linewidth]{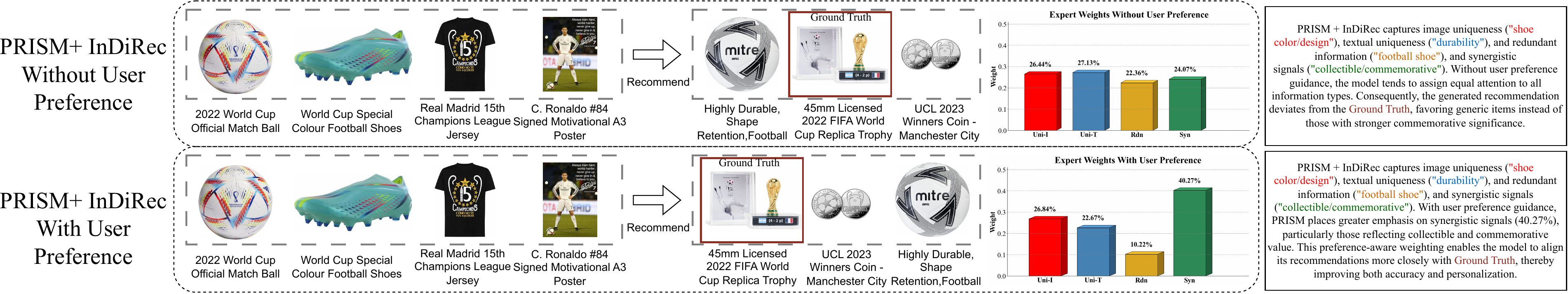}
    \caption{Case study of PRISM+InDiRec with and without user preference guidance. With preference guidance, the model increases emphasis on synergistic information, bringing the Ground Truth item to a higher rank.}
    \Description{Case study comparing REARM and PRISM+InDiRec.}
    \label{fig:case}
\end{figure*}

We take PRISM+InDiRec as a case study to illustrate how user preference guidance affects recommendation outcomes. As shown in Figure~\ref{fig:case}, the model effectively captures image uniqueness (e.g., shoe color/design, jersey appearance), textual uniqueness (e.g., durability, speed property), redundant information (e.g., football shoe, official jersey), and synergistic signals that convey collectible or commemorative value. These diverse cues contribute to a more comprehensive understanding of items. However, without user preference guidance, the model assigns nearly equal attention to all information types, leading to recommendations that deviate from the ground truth and prioritize generic items. In contrast, with user preference guidance, PRISM places greater emphasis on synergistic signals (40.27\%), particularly those reflecting collectible and commemorative value.  This adaptive weighting enables the model to better reproduce the ground-truth ranking, thereby generating more accurate and personalized recommendations.

\subsection{Visual Analysis (RQ4)}\label{sec:visual_analysis}

\begin{figure}[ht]
    \centering
    % 左图 REARM
    \begin{subfigure}[t]{0.48\linewidth}
        \centering
        \includegraphics[width=\linewidth]{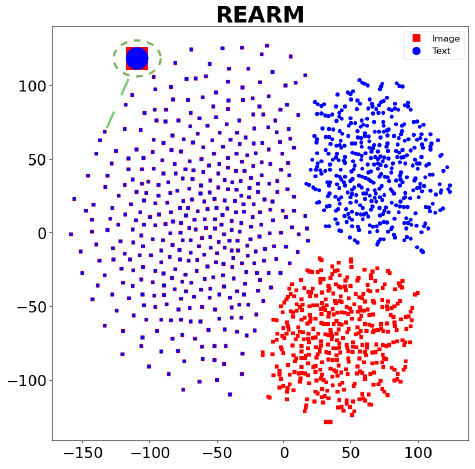}
        \caption{REARM.}
        \label{fig:rearm_tsne}
    \end{subfigure}
    \hfill
    % 右图 PRISM + InDiRec
    \begin{subfigure}[t]{0.48\linewidth}
        \centering
        \includegraphics[width=\linewidth]{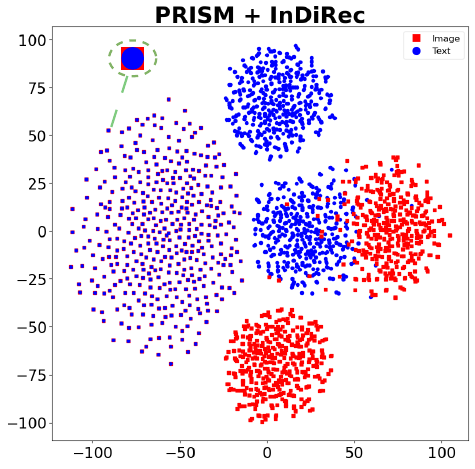}
        \caption{PRISM + InDiRec.}
        \label{fig:prism_tsne}
    \end{subfigure}
    \caption{The t-SNE visualization of item embeddings on the Yelp dataset.}
    \Description{The t-SNE visualization of item representations.}
    \label{fig:tsne_comparison}
\end{figure}

We employ t-SNE to visually verify PRISM can disentangle modality interactions into uniqueness, redundancy, and synergy, in contrast to REARM~\cite{REARM}, which fails to capture synergistic information. We randomly sample 200 items from the Yelp dataset and project their learned embeddings into a two-dimensional space, where blue circles denote text embeddings and red squares represent image embeddings. As shown in Figure~\ref{fig:rearm_tsne}, REARM produces three distinct clusters: a blue cluster representing text-unique information, a red cluster corresponding to image-unique information, and an overlapping mixed region where blue and red points coincide, indicating redundancy. Notably, no separate cluster corresponding to synergy is observed. In contrast, PRISM+InDiRec (Figure~\ref{fig:prism_tsne}) exhibits four clear clusters: a text-unique cluster (blue), an image-unique cluster (red), a redundancy cluster where text and image embeddings overlap, and an additional cluster where text and image embeddings are co-located but not overlapping. This new cluster represents synergistic information, defined as incremental signals that arise only from the interaction of image and textual uniqueness.

\subsection{Complexity Analysis (RQ5)}\label{sec:complexity}

We analyze the computational complexity of PRISM and show that its overhead is predictable and manageable. The additional cost mainly comes from the reweighting module and the interaction experts, both implemented as MLPs. (1) \textit{Time complexity:} The reweighting module incurs a cost of $\mathcal{O}((M+3) D H_{rw} + H_{rw} N_{exp})$, where $M$ is the number of modalities, $D$ is the embedding dimension, $H_{rw}$ is the hidden size of the reweighting module, and $N_{exp}$ is the number of experts. Each interaction expert is also an MLP with hidden size $H_e$, requiring $\mathcal{O}(D H_e + H_e D)$ per forward pass, which simplifies to $\mathcal{O}(D H_e)$. Since $(1+M)$ passes are needed to compute the full and masked representations for each expert, the total cost of all experts is $\mathcal{O}(N_{exp} (1+M) D H_e)$. Therefore, the overall time complexity of PRISM is $\mathcal{O}((M+3) D H_{rw} + H_{rw} N_{exp} + N_{exp} (1+M) D H_e)$. (2) \textit{Space complexity:} Parameter storage is $\mathcal{O}((M+3) D H_{rw} + H_{rw} N_{exp} + N_{exp} D H_e)$, and activation storage from $(1+M)$ forward passes is $\mathcal{O}(N_{exp} (1+M) B D)$, where $B$ is the batch size. Hence, the total space complexity is $\mathcal{O}((M+3) D H_{rw} + H_{rw} N_{exp} + N_{exp} D H_e + N_{exp} (1+M) B D)$. 

Both time and space complexity scale linearly with the number of experts $N_{exp}$, and activation storage further scales with $B$. Since $N_{exp}$ is small (four in PRISM), the overall computational cost remains comparable to standard multimodal fusion architectures. We empirically validate this analysis in Appendix~\ref{time_complexity}, where Table~\ref{tab:time} presents time and memory usage across datasets and backbones.
\section{Related Work}

\subsection{Sequential Recommendation}

Sequential recommendation (SR) aims to predict the next item a user will interact with based on their historical behavior sequence~\cite{SASRec, SR_Survey1, SR_Survey2}. Early approaches, including neighborhood-based models, factorization machines, and Markov chains~\cite{FM}, established the foundation for user–item interaction modeling but were limited in capturing complex and long-range dependencies. With the advent of deep learning, SR has evolved through diverse neural architectures, such as RNN-based (e.g., GRU4Rec~\cite{GRU4Rec}), CNN-based (e.g., Caser~\cite{Caser}), GNN-based (e.g., SURGE~\cite{SURGE}), and self-attention-based models (e.g., SASRec~\cite{SASRec}), which effectively capture both temporal dynamics and contextual correlations within sequences. Recently, contrastive and self-supervised frameworks~\cite{ICLRec, DuoRec} align augmented sequence views to improve generalization under sparse or noisy conditions. Nevertheless, most SR models remain unimodal, overlooking the rich multimodal content of items that could further refine user preference modeling.

\subsection{Multimodal Recommendation}

To improve recommendation quality, multimodal approaches leverage additional item modalities such as images and texts~\cite{VBPR, UniSRec, MMSBR, HM4SR}. Early studies typically relied on simple fusion; for example, VBPR~\cite{VBPR} directly concatenates visual features with ID embeddings without explicitly modeling inter-modal interactions. Later, self-supervised pretraining was introduced to enhance content-based representations (e.g., S3-Rec~\cite{S3Rec}).
As deep models matured, more sophisticated architectures emerged. UniSRec~\cite{UniSRec} leverages contrastive pretraining on textual content to enable semantic transfer; MMSR~\cite{MMSR} employs heterogeneous graphs to capture cross-modal dependencies; and TedRec~\cite{TedRec} explores frequency-domain integration to filter redundant signals across modalities. HM4SR~\cite{HM4SR} further introduces a hierarchical mixture-of-experts to adaptively fuse multimodal representations and capture evolving user interests.
Beyond fusion, MMSBR~\cite{MMSBR} introduces pseudo-modality contrastive learning for session-based recommendation. IISAN~\cite{IISAN} achieves efficient intra- and inter-modal adaptation through adapter-based parameter-efficient fine-tuning. Building on these advances, DGMRec~\cite{DGMRec} separates shared and modality-specific features using dedicated encoders and introduces a generative pathway to address missing modalities. REARM~\cite{REARM} refines multimodal contrastive learning by modeling higher-order user–item relations through graphs and employing a meta-network with orthogonal constraints to denoise shared features while preserving modality-specific signals.
However, by focusing primarily on shared and modality-specific representations without systematically modeling multimodal interactions, existing methods often overlook synergistic information, which is the emergent knowledge arising from the joint presence of multiple modalities, limiting their ability to adaptively fuse shared, specific, and synergistic signals.
\section{Conclusion}

In this paper, we proposed PRISM, a novel information-theoretic framework that addresses the challenges of fine-grained information disentanglement and user-adaptive fusion in MSR. PRISM is a plug-and-play module that decomposes multimodal information into unique, redundant, and synergistic components through an Interaction Expert Layer, and adaptively fuses them via an Adaptive Fusion Layer guided by user preferences. Extensive experiments demonstrate that PRISM consistently improves both performance and interpretability across diverse backbones, providing a unified and effective solution for multimodal fusion. Future work will explore extending PRISM to large-scale recommendation scenarios with diverse modality types and further enhancing adaptability by incorporating temporal dynamics of user preferences.

\section{Acknowledgments}
This work was supported by the Basic Research Program of Jiangsu (Grant No. BK20250668), Jiangsu Provincial Young Science and Technology Talent Support Program (Grants No. JSTJ-2025-944), Science and Technology Major Special Program of Jiangsu (Grants No. BG2024028), Basic Science (Natural Science) Research Project of Jiangsu Province Higher Education Institutions (Grants No. 25KJB-520039).

\bibliographystyle{ACM-Reference-Format}
\bibliography{WWW_2026}

@inproceedings{I2MoE,
  author       = {Jiayi Xin and
                  Sukwon Yun and
                  Jie Peng and
                  Inyoung Choi and
                  Jenna L. Ballard and
                  Tianlong Chen and
                  Qi Long},
  title        = {I2MoE: Interpretable Multimodal Interaction-aware Mixture-of-Experts},
  booktitle={International conference on machine learning},
  address = {USA},
  pages={},
  publisher  = {PMLR},
  year       = {2025},
}

@article{8269806,
    author = {Baltrusaitis, Tadas and Ahuja, Chaitanya and Morency, Louis-Philippe},
    title = {Multimodal Machine Learning: A Survey and Taxonomy},
    year = {2019},
    publisher = {IEEE Computer Society},
    address = {USA},
    volume = {41},
    number = {2},
    journal = {IEEE Trans. Pattern Anal. Mach. Intell.},
    pages = {423–443},
    numpages = {21}
}

@inproceedings{PAMD,
    author = {Han, Tengyue and Wang, Pengfei and Niu, Shaozhang and Li, Chenliang},
    title = {Modality Matches Modality: Pretraining Modality-Disentangled Item Representations for Recommendation},
    year = {2022},
    publisher = {Association for Computing Machinery},
    address = {New York, NY, USA},
    booktitle = {Proceedings of the ACM Web Conference 2022},
    pages = {2058–2066},
    numpages = {9},
    series = {WWW '22}
}

@inproceedings{yu2021,
  author       = {Wenmeng Yu and
                  Hua Xu and
                  Ziqi Yuan and
                  Jiele Wu},
  title        = {Learning Modality-Specific Representations with Self-Supervised Multi-Task
                  Learning for Multimodal Sentiment Analysis},
  booktitle    = {Proceedings of the Thirty-Fifth AAAI Conference on Artificial                     Intelligence},
  pages        = {10790--10797},
  publisher    = {{AAAI} Press},
  address      = {New York, NY, USA},
  year         = {2021},
}

@article{zhang2022,
    author = {Zhang, Jinghao and Zhu, Yanqiao and Liu, Qiang and Zhang, Mengqi and Wu, Shu and Wang, Liang},
    title = {Latent Structure Mining With Contrastive Modality Fusion for Multimedia Recommendation},
    year = {2023},
    publisher = {IEEE Educational Activities Department},
    address = {USA},
    volume = {35},
    number = {9},
    journal = {IEEE Trans. on Knowl. and Data Eng.},
    pages = {9154–9167},
    numpages = {14}
}

@article{liang2023,
    author = {Liang, Paul Pu and Zadeh, Amir and Morency, Louis-Philippe},
    title = {Foundations \& Trends in Multimodal Machine Learning: Principles, Challenges, and Open Questions},
    year = {2024},
    publisher = {Association for Computing Machinery},
    address = {New York, NY, USA},
    volume = {56},
    number = {10},
    journal = {ACM Comput. Surv.},
    numpages = {42},
}

@article{wollstadt2023,
    author = {Wollstadt, Patricia and Schmitt, Sebastian and Wibral, Michael},
    title = {A rigorous information-theoretic definition of redundancy and relevancy in feature selection based on (partial) information decomposition},
    year = {2023},
    publisher = {JMLR.org},
    volume = {24},
    number = {1},
    journal = {J. Mach. Learn. Res.},
    numpages = {44},
}

@inproceedings{liang2023quantifying,
    author = {Liang, Paul Pu and Cheng, Yun and Fan, Xiang and others},
    title = {Quantifying \& modeling multimodal interactions: an information decomposition framework},
    year = {2023},
    publisher = {Curran Associates Inc.},
    address = {Red Hook, NY, USA},
    booktitle = {Proceedings of the 37th International Conference on Neural Information Processing Systems},
    numpages = {43},
    series = {NIPS '23}
}

@inproceedings{SASRec,
  author       = {Wang{-}Cheng Kang and
                  Julian J. McAuley},
  title        = {Self-Attentive Sequential Recommendation},
  booktitle    = {{IEEE} International Conference on Data Mining, {ICDM} 2018, Singapore,
                  November 17-20, 2018},
  pages        = {197--206},
  publisher    = {{IEEE} Computer Society},
  year         = {2018},
  address     = {Singapore},
}

@inproceedings{Bert4rec,
author = {Sun, Fei and Liu, Jun and Wu, Jian and Pei, Changhua and Lin, Xiao and Ou, Wenwu and Jiang, Peng},
title = {BERT4Rec: Sequential Recommendation with Bidirectional Encoder Representations from Transformer},
year = {2019},
publisher = {Association for Computing Machinery},
address = {New York, NY, USA},
booktitle = {Proceedings of the 28th ACM International Conference on Information and Knowledge Management},
pages = {1441–1450},
numpages = {10},
series = {CIKM '19}
}

@inproceedings{UniSRec,
author = {Hou, Yupeng and Mu, Shanlei and Zhao, Wayne Xin and Li, Yaliang and Ding, Bolin and Wen, Ji-Rong},
title = {Towards Universal Sequence Representation Learning for Recommender Systems},
year = {2022},
publisher = {Association for Computing Machinery},
address = {New York, NY, USA},
booktitle = {Proceedings of the 28th ACM SIGKDD Conference on Knowledge Discovery and Data Mining},
pages = {585–593},
numpages = {9},
series = {KDD '22}
}

@inproceedings{VBPR,
author = {He, Ruining and McAuley, Julian},
title = {VBPR: visual Bayesian Personalized Ranking from implicit feedback},
year = {2016},
publisher = {AAAI Press},
booktitle = {Proceedings of the Thirtieth AAAI Conference on Artificial Intelligence},
pages = {144–150},
numpages = {7},
address = {Phoenix, Arizona},
series = {AAAI'16}
}

@inproceedings{MMMLP,
author = {Liang, Jiahao and Zhao, Xiangyu and Li, Muyang and Zhang, Zijian and Wang, Wanyu and Liu, Haochen and Liu, Zitao},
title={Mmmlp: Multi-modal multilayer perceptron for sequential recommendations},
year = {2023},
publisher = {Association for Computing Machinery},
address = {New York, NY, USA},
booktitle = {Proceedings of the ACM Web Conference 2023},
pages = {1109–1117},
numpages = {9},
series = {WWW '23}
}

@inproceedings{HM4SR,
author = {Zhang, Shengzhe and Chen, Liyi and Shen, Dazhong and Wang, Chao and Xiong, Hui},
title = {Hierarchical Time-Aware Mixture of Experts for Multi-Modal Sequential Recommendation},
year = {2025},
publisher = {Association for Computing Machinery},
address = {New York, NY, USA},
booktitle = {Proceedings of the ACM on Web Conference 2025},
pages = {3672–3682},
numpages = {11},
series = {WWW '25}
}

@article{SR_Survey1,
    author = {Jing, Mengyuan and Zhu, Yanmin and Zang, Tianzi and Wang, Ke},
    title = {Contrastive Self-supervised Learning in Recommender Systems: A Survey},
    year = {2023},
    publisher = {Association for Computing Machinery},
    address = {New York, NY, USA},
    volume = {42},
    number = {2},
    journal = {ACM Trans. Inf. Syst.},
    numpages = {39},
}

@article{SR_Survey2,
    author = {Fang, Hui and Zhang, Danning and Shu, Yiheng and Guo, Guibing},
    title = {Deep Learning for Sequential Recommendation: Algorithms, Influential Factors, and Evaluations},
    year = {2020},
    publisher = {Association for Computing Machinery},
    address = {New York, NY, USA},
    volume = {39},
    number = {1},
    journal = {ACM Trans. Inf. Syst.},
    numpages = {42},
}

@inproceedings{TrustSVD,
    author = {Guo, Guibing and Zhang, Jie and Yorke-Smith, Neil},
    title = {TrustSVD: collaborative filtering with both the explicit and implicit influence of user trust and of item ratings},
    year = {2015},
    publisher = {AAAI Press},
    booktitle = {Proceedings of the Twenty-Ninth AAAI Conference on Artificial Intelligence},
    address = {New York, NY, USA},
    pages = {123–129},
    numpages = {7},
    series = {AAAI'15}
}

@inproceedings{AlignRec,
    author = {Liu, Yifan and Zhang, Kangning and Ren, Xiangyuan and Huang, Yanhua and Jin, Jiarui and Qin, Yingjie and Su, Ruilong and Xu, Ruiwen and Yu, Yong and Zhang, Weinan},
    title = {AlignRec: Aligning and Training in Multimodal Recommendations},
    year = {2024},
    publisher = {Association for Computing Machinery},
    address = {New York, NY, USA},
    booktitle = {Proceedings of the 33rd ACM International Conference on Information and Knowledge Management},
    pages = {1503–1512},
    numpages = {10},
    series = {CIKM '24}
}

@inproceedings{Mentor,
    author = {Xu, Jinfeng and Chen, Zheyu and Yang, Shuo and Li, Jinze and Wang, Hewei and Ngai, Edith C. H.},
    title = {MENTOR: multi-level self-supervised learning for multimodal recommendation},
    year = {2025},
    publisher = {AAAI Press},
    booktitle = {Proceedings of the Thirty-Ninth AAAI Conference on Artificial Intelligence},
    address = {New York, NY, USA},
    numpages = {10},
    series = {AAAI'25}
}

@ARTICLE{MMSBR,
  author={Zhang, Xiaokun and Xu, Bo and Ma, Fenglong and Li, Chenliang and Yang, Liang and Lin, Hongfei},
  journal={IEEE Transactions on Knowledge and Data Engineering}, 
  title={Beyond Co-Occurrence: Multi-Modal Session-Based Recommendation}, 
  year={2024},
  volume={36},
  number={4},
  pages={1450-1462},
}

@inproceedings{MMSSL,
    author = {Wei, Wei and Huang, Chao and Xia, Lianghao and Zhang, Chuxu},
    title = {Multi-Modal Self-Supervised Learning for Recommendation},
    year = {2023},
    publisher = {Association for Computing Machinery},
    address = {New York, NY, USA},
    booktitle = {Proceedings of the ACM Web Conference 2023},
    pages = {790–800},
    numpages = {11},
    series = {WWW '23}
}

@misc{SEA,
      title={It is Never Too Late to Mend: Separate Learning for Multimedia Recommendation}, 
      author={Zhuangzhuang He and Zihan Wang and Yonghui Yang and Haoyue Bai and Le Wu},
      year={2024},
      eprint={2406.08270},
      archivePrefix={arXiv},
      primaryClass={cs.IR},
}

@inproceedings{DGMRec,
    author = {Kim, Jiwan and Kang, Hongseok and Kim, Sein and Kim, Kibum and Park, Chanyoung},
    title = {Disentangling and Generating Modalities for Recommendation in Missing Modality Scenarios},
    year = {2025},
    publisher = {Association for Computing Machinery},
    address = {New York, NY, USA}, 
    booktitle = {Proceedings of the 48th International ACM SIGIR Conference on Research and Development in Information Retrieval},
    pages = {1820–1829},
    numpages = {10},
    series = {SIGIR '25}
}

@inproceedings{MMSR,
    author = {Hu, Hengchang and Guo, Wei and Liu, Yong and Kan, Min-Yen},
    title = {Adaptive Multi-Modalities Fusion in Sequential Recommendation Systems},
    year = {2023},
    publisher = {Association for Computing Machinery},
    address = {New York, NY, USA},
    booktitle = {Proceedings of the 32nd ACM International Conference on Information and Knowledge Management},
    pages = {843–853},
    numpages = {11},
    series = {CIKM '23}
}

@inproceedings{TedRec,
    author = {Xu, Lanling and Tian, Zhen and Li, Bingqian and others},
    title = {Sequence-level Semantic Representation Fusion for Recommender Systems},
    year = {2024},
    publisher = {Association for Computing Machinery},
    address = {New York, NY, USA},
    booktitle = {Proceedings of the 33rd ACM International Conference on Information and Knowledge Management},
    pages = {5015–5022},
    numpages = {8},
    series = {CIKM '24}
}

@inproceedings{STOSA,
    author = {Fan, Ziwei and Liu, Zhiwei and Wang, Yu and Wang, Alice and Nazari, Zahra and Zheng, Lei and Peng, Hao and Yu, Philip S.},
    title = {Sequential Recommendation via Stochastic Self-Attention},
    year = {2022},
    publisher = {Association for Computing Machinery},
    address = {New York, NY, USA},
    booktitle = {Proceedings of the ACM Web Conference 2022},
    pages = {2036–2047},
    numpages = {12},
    series = {WWW '22}
}

@inproceedings{whattoalign,
    title     = {What to align in multimodal contrastive learning?},
    author    = {Benoit Dufumier and Javiera Castillo Navarro and Devis Tuia and Jean-Philippe Thiran},
    booktitle = {The Thirteenth International Conference on Learning Representations (ICLR)},
    year      = {2025},
    publisher = {ICLR},
    address   = {Singapore},
    pages     = {},
}

@article{MLP,
    title={The perceptron: a probabilistic model for information storage and organization in the brain.},
    author={Rosenblatt, Frank},
    journal={Psychological review},
    volume={65},
    number={6},
    pages={386},
    year={1958},
    publisher={American Psychological Association}
}

@inproceedings{Transformer,
    author = {Vaswani, Ashish and Shazeer, Noam and Parmar, Niki and Uszkoreit, Jakob and Jones, Llion and Gomez, Aidan N and Kaiser, \L ukasz and Polosukhin, Illia},
    booktitle = {Advances in Neural Information Processing Systems},
    pages = {},
    publisher = {Curran Associates, Inc.},
    title = {Attention is All you Need},
    volume = {30},
    year = {2017},
    address={San Diego, CA, USA},
}

@inproceedings{BPR,
    author = {Rendle, Steffen and Freudenthaler, Christoph and Gantner, Zeno and Schmidt-Thieme, Lars},
    title = {BPR: Bayesian personalized ranking from implicit feedback},
    year = {2009},
    publisher = {AUAI Press},
    address = {Arlington, Virginia, USA},
    booktitle = {Proceedings of the Twenty-Fifth Conference on Uncertainty in Artificial Intelligence},
    pages = {452–461},
    numpages = {10},
    series = {UAI '09}
}

@article{Diffurec_TOIS,
    author = {Li, Zihao and Sun, Aixin and Li, Chenliang},
    title = {DiffuRec: A Diffusion Model for Sequential Recommendation},
    year = {2023},
    publisher = {Association for Computing Machinery},
    address = {New York, NY, USA},
    volume = {42},
    number = {3},
    journal = {ACM Trans. Inf. Syst.},
    numpages = {28},
}

@inproceedings{LightGCN,
    author = {He, Xiangnan and Deng, Kuan and Wang, Xiang and Li, Yan and Zhang, YongDong and Wang, Meng},
    title = {LightGCN: Simplifying and Powering Graph Convolution Network for Recommendation},
    year = {2020},
    publisher = {Association for Computing Machinery},
    address = {New York, NY, USA},
    booktitle = {Proceedings of the 43rd International ACM SIGIR Conference on Research and Development in Information Retrieval},
    pages = {639–648},
    numpages = {10},
    series = {SIGIR '20}
}

@inproceedings{Yelp1,
    author = {Zhang, Xiaokun and Xu, Bo and Wu, Youlin and Zhong, Yuan and Lin, Hongfei and Ma, Fenglong},
    title = {FineRec: Exploring Fine-grained Sequential Recommendation},
    year = {2024},
    publisher = {Association for Computing Machinery},
    address = {New York, NY, USA},
    booktitle = {Proceedings of the 47th International ACM SIGIR Conference on Research and Development in Information Retrieval},
    pages = {1599–1608},
    numpages = {10},
    series = {SIGIR '24}
}

@inproceedings{Yelp2,
    author = {Li, Xuewei and Sun, Aitong and Zhao, Mankun and Yu, Jian and Zhu, Kun and Jin, Di and Yu, Mei and Yu, Ruiguo},
    title = {Multi-Intention Oriented Contrastive Learning for Sequential Recommendation},
    year = {2023},
    publisher = {Association for Computing Machinery},
    address = {New York, NY, USA},
    booktitle = {Proceedings of the Sixteenth ACM International Conference on Web Search and Data Mining},
    pages = {411–419},
    numpages = {9},
    series = {WSDM '23}
}

@inproceedings{tripletloss,
  author       = {Florian Schroff and
                  Dmitry Kalenichenko and
                  James Philbin},
  title        = {FaceNet: {A} unified embedding for face recognition and clustering},
  booktitle    = {{IEEE} Conference on Computer Vision and Pattern Recognition, {CVPR}
                  2015, Boston, MA, USA, June 7-12, 2015},
  pages        = {815--823},
  publisher    = {{IEEE} Computer Society},
  address      = {USA},
  year         = {2015},
}

@inproceedings{Cossim,
    author = {Nguyen, Hieu V. and Bai, Li},
    title = {Cosine similarity metric learning for face verification},
    year = {2010},
    publisher = {Springer-Verlag},
    address = {Berlin, Heidelberg},
    booktitle = {Proceedings of the 10th Asian Conference on Computer Vision - Volume Part II},
    pages = {709–720},
    numpages = {12},
    series = {ACCV'10}
}

@inproceedings{InDiRec,
    author = {Qu, Yuanpeng and Nobuhara, Hajime},
    title = {Intent-aware Diffusion with Contrastive Learning for Sequential Recommendation},
    year = {2025},
    publisher = {Association for Computing Machinery},
    address = {New York, NY, USA},
    booktitle = {Proceedings of the 48th International ACM SIGIR Conference on Research and Development in Information Retrieval},
    pages = {1552–1561},
    numpages = {10},
    series = {SIGIR '25}
}

@inproceedings{FM,
    author = {Rendle, Steffen and Freudenthaler, Christoph and Schmidt-Thieme, Lars},
    title = {Factorizing personalized Markov chains for next-basket recommendation},
    year = {2010},
    publisher = {Association for Computing Machinery},
    address = {New York, NY, USA},
    booktitle = {Proceedings of the 19th International Conference on World Wide Web},
    pages = {811–820},
    numpages = {10},
    series = {WWW '10}
}

@inproceedings{Caser,
    author = {Tang, Jiaxi and Wang, Ke},
    title = {Personalized Top-N Sequential Recommendation via Convolutional Sequence Embedding},
    year = {2018},
    publisher = {Association for Computing Machinery},
    address = {New York, NY, USA},
    booktitle = {Proceedings of the Eleventh ACM International Conference on Web Search and Data Mining},
    pages = {565–573},
    numpages = {9},
    series = {WSDM '18}
}

@inproceedings{IISAN,
    author = {Fu, Junchen and Ge, Xuri and Xin, Xin and Karatzoglou, Alexandros and Arapakis, Ioannis and Wang, Jie and Jose, Joemon M.},
    title = {IISAN: Efficiently Adapting Multimodal Representation for Sequential Recommendation with Decoupled PEFT},
    year = {2024},
    publisher = {Association for Computing Machinery},
    address = {New York, NY, USA},
    booktitle = {Proceedings of the 47th International ACM SIGIR Conference on Research and Development in Information Retrieval},
    pages = {687–697},
    numpages = {11},
    series = {SIGIR '24}
}

@inproceedings{S3Rec,
    author = {Zhou, Kun and Wang, Hui and Zhao, Wayne Xin and Zhu, Yutao and Wang, Sirui and Zhang, Fuzheng and Wang, Zhongyuan and Wen, Ji-Rong},
    title = {S3-Rec: Self-Supervised Learning for Sequential Recommendation with Mutual Information Maximization},
    year = {2020},
    publisher = {Association for Computing Machinery},
    address = {New York, NY, USA},
    booktitle = {Proceedings of the 29th ACM International Conference on Information \& Knowledge Management},
    pages = {1893–1902},
    numpages = {10},
    series = {CIKM '20}
}

@inproceedings{GRU4Rec,
  author       = {Bal{\'{a}}zs Hidasi and
                  Alexandros Karatzoglou and
                  Linas Baltrunas and
                  Domonkos Tikk},
  editor       = {Yoshua Bengio and
                  Yann LeCun},
  title        = {Session-based Recommendations with Recurrent Neural Networks},
  booktitle    = {4th International Conference on Learning Representations, {ICLR} 2016,
                  San Juan, Puerto Rico, May 2-4, 2016, Conference Track Proceedings},
  year         = {2016},
  publisher = {ICLR},            
  address   = {San Juan, Puerto Rico},  
  pages    = {} 
}

@inproceedings{REARM,
  title     = {Refining Contrastive Learning and Homography Relations for Multi-Modal Recommendation},
  author    = {Ma, Shouxing and 
               Zeng, Yawen and 
               Wu, Shiqing and 
               Xu, Guandong},
  booktitle = {Proceedings of the 33th ACM International Conference on Multimedia},
  year      = {2025},
  address   = {Dublin, Ireland},
  publisher = {Association for Computing Machinery},
  pages    = {} 
}

@article{liu2022TMM_disentangled,
    author = {Liu, Fan and Chen, Huilin and Cheng, Zhiyong and Liu, Anan and Nie, Liqiang and Kankanhalli, Mohan},
    title = {Disentangled Multimodal Representation Learning for Recommendation},
    year = {2023},
    publisher = {IEEE Press},
    journal = {Trans. Multi.},
    volume = {25},
    numpages = {11},
    pages = {7149–7159},
}

@misc{xu2025survey,
      title={A Survey on Multimodal Recommender Systems: Recent Advances and Future Directions}, 
      author={Jinfeng Xu and Zheyu Chen and Shuo Yang and Jinze Li and Wei Wang and Xiping Hu and Steven Hoi and Edith Ngai},
      year={2025},
      eprint={2502.15711},
      archivePrefix={arXiv},
      primaryClass={cs.IR},
      url={https://arxiv.org/abs/2502.15711}, 
}

@inproceedings{Recbole,
    author = {Zhao, Wayne Xin and Hou, Yupeng and others},
    title = {RecBole 2.0: Towards a More Up-to-Date Recommendation Library},
    year = {2022},
    publisher = {Association for Computing Machinery},
    address = {New York, NY, USA},
    booktitle = {Proceedings of the 31st ACM International Conference on Information \& Knowledge Management},
    pages = {4722–4726},
    numpages = {5},
    series = {CIKM '22}
}

@inproceedings{SURGE,
    author = {Chang, Jianxin and Gao, Chen and Zheng, Yu and Hui, Yiqun and Niu, Yanan and Song, Yang and Jin, Depeng and Li, Yong},
    title = {Sequential Recommendation with Graph Neural Networks},
    year = {2021},
    publisher = {Association for Computing Machinery},
    address = {New York, NY, USA},
    booktitle = {Proceedings of the 44th International ACM SIGIR Conference on Research and Development in Information Retrieval},
    pages = {378–387},
    numpages = {10},
    series = {SIGIR '21}
}

@inproceedings{ICLRec,
    author = {Chen, Yongjun and Liu, Zhiwei and Li, Jia and McAuley, Julian and Xiong, Caiming},
    title = {Intent Contrastive Learning for Sequential Recommendation},
    year = {2022},
    publisher = {Association for Computing Machinery},
    address = {New York, NY, USA},
    booktitle = {Proceedings of the ACM Web Conference 2022},
    pages = {2172–2182},
    numpages = {11},
    series = {WWW '22}
}

@inproceedings{DuoRec,
    author = {Qiu, Ruihong and Huang, Zi and Yin, Hongzhi and Wang, Zijian},
    title = {Contrastive Learning for Representation Degeneration Problem in Sequential Recommendation},
    year = {2022},
    publisher = {Association for Computing Machinery},
    address = {New York, NY, USA},
    booktitle = {Proceedings of the Fifteenth ACM International Conference on Web Search and Data Mining},
    pages = {813–823},
    numpages = {11},
    series = {WSDM '22}
}

@inproceedings{yu-etal-2024-mmoe,
    title = "{MM}o{E}: Enhancing Multimodal Models with Mixtures of Multimodal Interaction Experts",
    author = "Yu, Haofei  and
      Qi, Zhengyang  and
      Jang, Lawrence Keunho  and
      Salakhutdinov, Russ  and
      Morency, Louis-Philippe  and
      Liang, Paul Pu",
    editor = "Al-Onaizan, Yaser  and
      Bansal, Mohit  and
      Chen, Yun-Nung",
    booktitle = "Proceedings of the 2024 Conference on Empirical Methods in Natural Language Processing",
    month = nov,
    year = "2024",
    address = "Miami, Florida, USA",
    publisher = "Association for Computational Linguistics",
    url = "https://aclanthology.org/2024.emnlp-main.558/",
    doi = "10.18653/v1/2024.emnlp-main.558",
    pages = "10006--10030",
}

@inproceedings{BCE,
author = {Zhang, Zhilu and Sabuncu, Mert R.},
title = {Generalized cross entropy loss for training deep neural networks with noisy labels},
year = {2018},
publisher = {Curran Associates Inc.},
address = {Red Hook, NY, USA},
booktitle = {Proceedings of the 32nd International Conference on Neural Information Processing Systems},
pages = {8792–8802},
numpages = {11},
series = {NIPS'18}
}

@inproceedings{MGCN,
author = {Yu, Penghang and Tan, Zhiyi and Lu, Guanming and Bao, Bing-Kun},
title = {Multi-View Graph Convolutional Network for Multimedia Recommendation},
year = {2023},
publisher = {Association for Computing Machinery},
address = {New York, NY, USA},
booktitle = {Proceedings of the 31st ACM International Conference on Multimedia},
pages = {6576–6585},
numpages = {10},
series = {MM '23}
}

@inproceedings{10.1145/3219819.3220007,
author = {Ma, Jiaqi and Zhao, Zhe and Yi, Xinyang and Chen, Jilin and Hong, Lichan and Chi, Ed H.},
title = {Modeling Task Relationships in Multi-task Learning with Multi-gate Mixture-of-Experts},
year = {2018},
publisher = {Association for Computing Machinery},
address = {New York, NY, USA},
booktitle = {Proceedings of the 24th ACM SIGKDD International Conference on Knowledge Discovery \& Data Mining},
pages = {1930–1939},
numpages = {10},
series = {KDD '18}
}

@article{Cai_2025,
  author={Cai, Weilin and Jiang, Juyong and Wang, Fan and Tang, Jing and Kim, Sunghun and Huang, Jiayi},
  journal={IEEE Transactions on Knowledge and Data Engineering}, 
  title={A Survey on Mixture of Experts in Large Language Models}, 
  year={2025},
  volume={37},
  number={7},
  pages={3896-3915},
}

@inproceedings{M3SRec,
author = {Bian, Shuqing and Pan, Xingyu and Zhao, Wayne Xin and Wang, Jinpeng and Wang, Chuyuan and Wen, Ji-Rong},
title = {Multi-modal Mixture of Experts Represetation Learning for Sequential Recommendation},
year = {2023},
publisher = {Association for Computing Machinery},
address = {New York, NY, USA},
booktitle = {Proceedings of the 32nd ACM International Conference on Information and Knowledge Management},
pages = {110–119},
numpages = {10},
series = {CIKM '23}
}
\appendix
\section{Appendix}

\subsection{Algorithm}\label{appendix:Algorithm}

Algorithm~\ref{alg:prism} details how PRISM plugs into a standard SR backbone.
\begin{algorithm}[!ht]
\caption{Sequential Recommendation with PRISM.}
\label{alg:prism}
\begin{algorithmic}[1]
\Require The interaction sequence of users $\mathcal{S} = \{x_1, x_2, \dots, x_n\}$
\Require ID embeddings layer $\text{itemEmb}(\cdot)$, modality-specific encoders $\text{imgEmb}(\cdot)$ and $\text{textEmb}(\cdot)$
\Require Interaction experts $\{E_j\}_{j \in \{\text{uni-i}, \text{uni-t}, \text{syn}, \text{rdn}\}}$
\Require Reweighting module $\mathrm{W}$, and interaction loss functions $\{\mathcal{L}_j\}$
\Require $\text{SE}$: Sequence Encoder, $\text{Pred}$: Prediction Layer 

\Procedure{SequentialRec\_with\_PRISM}{$\mathcal{S}$}
    \State \textit{// 1. Encode ID and multimodal features}
    \State $\mathbf{e}^{img} \gets \text{imgEmb}(x^{img}), \quad \mathbf{e}^{txt} \gets \text{textEmb}(x^{txt})$
    \State $\mathbf{e}^{id} \gets \text{itemEmb}(x^{id})$
    \Statex

    \State \textit{// 2. Expert modeling with masked modality inputs (only used during training)}
    \For{each expert $\{E_j\}_{j \in \{\text{uni-i}, \text{uni-t}, \text{syn}, \text{rdn}\}}$}
        \State $y \gets \text{Pred}(SE(E_j(\mathbf{e}^{img}, \mathbf{e}^{txt}),\, \mathbf{e}^{id}))$ \Comment{Full input}
        \State $y^{img} \gets \text{Pred}(SE(E_j(\mathbf{r}, \mathbf{e}^{txt}),\,\mathbf{e}^{id}))$ \Comment{Masked image}
        \State $y^{txt} \gets \text{Pred}(SE(E_j(\mathbf{e}^{img}, \mathbf{r}),\,\mathbf{e}^{id}))$ \Comment{Masked text}
    \EndFor
    \Statex
    \State \textit{// 3. Compute interaction losses (only used during training experts)}
    \State $\mathcal{L}_{\text{uni-i}} \gets \text{Uni-I Loss}(y, y^{img}, y^{txt})$
    \State $\mathcal{L}_{\text{uni-t}} \gets \text{Uni-T Loss}(y, y^{txt}, y^{img})$
    \State $\mathcal{L}_{\text{syn}} \gets \text{Syn Loss}(y, y^{img}, y^{txt})$
    \State $\mathcal{L}_{\text{rdn}} \gets \text{Rdn Loss}(y, y^{img}, y^{txt})$
    \State $\mathcal{L}_{\text{exp}} \gets \sum_{j \in \{\text{uni-i},\, \text{uni-t},\, \text{syn},\, \text{rdn}\}} \lambda_j \mathcal{L}_j$
    \Statex
    \State \textit{// 4. Compute total loss (only used during training $\mathrm{W}$)}
    \State $\mathcal{L} \gets \mathcal{L}_{\text{exp}} + \mathcal{L}_{\text{rec}}$
    \Statex
    \State \textit{// 5. Experts modeling interactions}
    \State $\mathbf{e}^j \gets E_j(\mathbf{e}^{img}, \mathbf{e}^{txt})$ \Comment{$j \in \{\text{uni-i},\, \text{uni-t},\, \text{syn},\, \text{rdn}\}$}
    \Statex
    \State \textit{// 6. Adaptive multimodal fusion}
    \State $w^j \gets \mathrm{W}(\mathbf{e}^j, \mathbf{e}^{id})$ \Comment{$j \in \{\text{uni-i},\, \text{uni-t},\, \text{syn},\, \text{rdn}\}$}
    \State $\mathbf{e}^m \gets \sum_{j} w^j \cdot \mathbf{e}^j$ \Comment{$j \in \{\text{uni-i},\, \text{uni-t},\, \text{syn},\, \text{rdn}\}$}
    \Statex
    \State \textit{// 7. Form final user representation and generate prediction}
    \State $\mathbf{h} \gets \text{SE}(\mathbf{e}^{id}, \mathbf{e}^{m})$
    \State $\mathbf{y} \gets \text{Pred}(\mathbf{h})$
    \State \Return $(\mathbf{y})$
\EndProcedure
\end{algorithmic}
\end{algorithm}

\subsection{Random Vector Justification}\label{appendix:random-vector-justification}

To empirically validate our masking design, we conducted an ablation study that compared three common strategies, namely random, mean, and zero-vector replacements, across four benchmark datasets (Home, Beauty, Sports, and Yelp). As summarized in Table~\ref{table:random-vector-multidata}, random vector masking consistently delivers the strongest performance on most metrics. For example, in the Beauty and Yelp datasets, the random strategy yields clear improvements in both Recall and NDCG over mean and zero replacements. While mean masking occasionally achieves competitive results on isolated metrics (e.g., N@20 on Sports), these gains are neither stable nor consistent across datasets. Empirical evidence demonstrates that the random masking strategy achieves stable superiority across different multimodal recommendation scenarios, thereby validating its selection as the mechanism in our framework.

\begin{table}[!ht]
    \caption{Performance comparison across different modality masking strategies (Random, Mean, Zero) on four datasets using the PRISM+InDiRec model. All results are averaged over 5 runs to ensure robustness and are statistically significant with \(p < 0.05\).}
    \label{table:random-vector-multidata}
    \centering
    \begin{tabular}{l l cccc}
        \toprule
        \textbf{Dataset} & \textbf{Strategy} & \textbf{R@10} & \textbf{R@20} & \textbf{N@10} & \textbf{N@20} \\
        \midrule
        \multirow{3}{*}{Home} 
            & Random & \textbf{0.0364} & \textbf{0.0501} & 0.0235 & \textbf{0.0272} \\
            & Mean   & 0.0358 & 0.0495 & \textbf{0.0240} & 0.0267 \\
            & Zero   & 0.0349 & 0.0483 & 0.0227 & 0.0262 \\
        \midrule
        \multirow{3}{*}{Beauty} 
            & Random & \textbf{0.1012} & \textbf{0.1388} & \textbf{0.0618} & \textbf{0.0711} \\
            & Mean   & 0.0997 & 0.1375 & 0.0602 & 0.0705 \\
            & Zero   & 0.1005 & 0.1369 & 0.0598 & 0.0690 \\ 
        \midrule
        \multirow{3}{*}{Sports} 
            & Random & \textbf{0.0557} & \textbf{0.0809} & \textbf{0.0337} & 0.0401 \\
            & Mean   & 0.0548 & 0.0802 & 0.0329 & \textbf{0.0403} \\ 
            & Zero   & 0.0532 & 0.0785 & 0.0318 & 0.0390 \\
        \midrule
        \multirow{3}{*}{Yelp} 
            & Random & \textbf{0.0422} & \textbf{0.0622} & \textbf{0.0305} & \textbf{0.0352} \\
            & Mean   & 0.0376 & 0.0582 & 0.0271 & 0.0320 \\
            & Zero   & 0.0361 & 0.0572 & 0.0257 & 0.0302 \\
        \bottomrule
    \end{tabular}
\end{table}

\begin{table*}[!ht]
\centering
\caption{Efficiency analysis of PRISM-enhanced models. We report the average training time per epoch and peak GPU memory consumption for SASRec, STOSA, and InDiRec, both with and without PRISM.
\textit{w/o} denotes the model \textit{without} PRISM, and \textit{w/} denotes the model \textit{with} PRISM. All results are averaged over 5 runs to ensure robustness and are statistically significant with \(p < 0.05\).}
\setlength\tabcolsep{5pt}
\label{tab:efficiency}
\begin{tabular}{cc|cc|cc|cc}
\toprule
\multirow{2}{*}{Dataset} & \multirow{2}{*}{Metric} 
& \multicolumn{2}{c|}{SASRec} 
& \multicolumn{2}{c|}{STOSA} 
& \multicolumn{2}{c}{InDiRec} \\
\cmidrule(lr){3-8}
& & \textit{w/o} PRISM & \textit{w}/ PRISM & \textit{w/o} PRISM & \textit{w/} PRISM & \textit{w/o} PRISM & \textit{w/} PRISM \\
\midrule
\multirow{2}{*}{Home} 
& Time (s/epoch) & 20.35 & 30.11 & 26.99 & 40.98 & 201.41 & 220.82 \\
& Memory (GB)    & 1.16  & 1.75  & 2.37  & 2.75  & 3.77 & 4.54 \\
\midrule
\multirow{2}{*}{Beauty} 
& Time (s/epoch) & 5.75  & 9.67  & 7.40  & 11.88 & 36.29 & 40.55 \\
& Memory (GB)    & 1.14  & 1.41  & 2.27  & 2.60  & 3.40 & 4.56 \\
\midrule
\multirow{2}{*}{Sports} 
& Time (s/epoch) & 7.19  & 20.81 & 17.11 & 28.81 & 123.51 & 141.44 \\
& Memory (GB)    & 1.33  & 1.51  & 2.33  & 2.66  & 3.55 & 4.12 \\
\midrule
\multirow{2}{*}{Yelp} 
& Time (s/epoch) & 742.67 & 799.44 & 778.21 & 834.69 & 1375.49 & 1439.11 \\
& Memory (GB)    & 1.45  & 4.61  & 2.56  & 7.51  & 3.96 & 10.47 \\
\bottomrule
\label{tab:time}
\end{tabular}
\end{table*}

\subsection{Theoretical Connection between Interaction Loss and PID} \label{appendix:connection-loss-and-pid}

Partial Information Decomposition (PID)~\cite{wollstadt2023, liang2023quantifying, 8269806, liang2023} provides a principled framework to separate the contribution of two sources $X^{img}$ and $X^{txt}$ (corresponding to $\mathbf{e}^{img}$ and $\mathbf{e}^{txt}$) to a target $T$. Formally, the mutual information can be decomposed into four non-overlapping components:

\begin{equation}
\begin{aligned}
I(T; X^{img}, X^{txt}) &= \mathrm{Red}(T; X^{img}, X^{txt}) + \mathrm{Syn}(T; X^{img}, X^{txt}) \\
&+ \mathrm{Unq}(T; X^{img} \setminus X^{txt}) + \mathrm{Unq}(T; X^{txt} \setminus X^{img}).\\
\end{aligned}    
\end{equation}

In our framework, the Interaction Expert Layer implements this decomposition by training four experts with perturbed inputs, such that each objective approximates one PID term.

\textbf{Uniqueness.}
When one modality is preserved while the other is replaced by a random vector $\mathbf{r}$, the corresponding expert prediction is forced to rely solely on the preserved modality. Specifically, the image uniqueness expert $E_{\text{uni-i}}$ compares the full prediction $y$ with the image-only prediction $y^{img}$, while penalizing similarity to the text-only prediction $y^{txt}$:
\begin{equation}
\mathcal{L}_{\text{uni-i}} = \mathrm{Triplet}(y, y^{img}, y^{txt}).
\end{equation}
Under the assumption that $\mathbf{r}$ contains no task-relevant information, this loss enforces specialization toward
\begin{equation}
\mathcal{L}_{\text{uni-i}} \;\propto\; \mathrm{Unq}(T; X^{img} \setminus X^{txt}).
\end{equation}

In a symmetric manner, the text uniqueness expert $E_{\text{uni-t}}$ operates with the roles reversed, where the text-only prediction $y^{txt}$ is treated as positive and the image-only prediction $y^{img}$ as negative, thereby approximating
\begin{equation}
\mathcal{L}_{\text{uni-t}} \;\propto\; \mathrm{Unq}(T; X^{txt} \setminus X^{img}).
\end{equation}

\textbf{Redundancy.}
For information redundantly encoded in both modalities, the model prediction should remain stable even when one modality is masked. The redundancy expert therefore maximizes the consistency between $y$, $y^{img}$, and $y^{txt}$:
\begin{equation}
\mathcal{L}_{\text{rdn}} = 1 - \tfrac{1}{2}\Big[ \mathrm{CosSim}(y,y^{img}) + \mathrm{CosSim}(y,y^{txt})\Big].
\end{equation}
This objective encourages the expert to capture modality-invariant patterns, thereby approximating
\begin{equation}
\mathcal{L}_{\text{rdn}} \;\propto\; \mathrm{Red}(T; X^{img}, X^{txt}).
\end{equation}

\textbf{Synergy.}
Synergistic information emerges only when both moda-lities are jointly observed, and disappears when either modality is masked. To isolate this effect, the synergy expert penalizes agreement between $y$ and the masked predictions:
\begin{equation}
\mathcal{L}_{\text{syn}} = \tfrac{1}{2}\Big[ \mathrm{CosSim}(y,y^{img}) + \mathrm{CosSim}(y,y^{txt})\Big].
\end{equation}
This design ensures that $E_{\text{syn}}$ emphasizes complementary cross-modal signals, approximating
\begin{equation}
\mathcal{L}_{\text{syn}} \;\propto\; \mathrm{Syn}(T; X^{img}, X^{txt}).
\end{equation}

\subsection{Time and Space Complexity}\label{time_complexity}

We analyze the computational overhead introduced by integrating PRISM into different SR backbones, as reported in Table~\ref{tab:time}. Across all datasets, PRISM consistently increases training time and GPU memory consumption due to the additional computation from its interaction experts and reweighting module. On the Beauty dataset, the per-epoch training time increases moderately, while on Sports, the overhead varies with backbone complexity. Lighter models such as SASRec experience a more noticeable slowdown compared to the heavier InDiRec. On the large-scale Yelp dataset, the additional cost remains marginal since attention operations dominate the runtime, demonstrating PRISM’s scalability.

In terms of memory, PRISM introduces a moderate increase caused by the extra parameters of the experts and multimodal activations, with small growth on medium-scale datasets and more visible growth on Yelp. Overall, PRISM adds stable and predictable overhead that remains within a manageable range. The total runtime is mainly determined by the underlying SR backbone, and the relative overhead decreases as the model size or dataset scale grows, making the trade-off acceptable given the consistent performance improvements.

\end{document}